\documentclass[5p]{elsarticle}

\usepackage{tikz}
\usepackage{xspace}
\usepackage{verbatim}
\usepackage{cprotect}
\usepackage{amsmath,amssymb}
\usepackage{tabularx}
\usepackage{todonotes}
\usepackage{multirow}
\usepackage{subfigure}
\usepackage{listings}
\usepackage{hyperref}
\usepackage{algorithm}
\usepackage{algpseudocode}

\newcolumntype{L}{>{\raggedright\arraybackslash}X}
\newcolumntype{R}{>{\raggedleft\arraybackslash}X}
\newcolumntype{Y}{>{\centering\arraybackslash}X}
\newcolumntype{C}{>{\Centering\arraybackslash}X}

\newcommand{\argmax}{\operatornamewithlimits{argmax}}

\definecolor{mygray}{rgb}{0.8,0.8,0.8}
\def\linear{\textit{LINEAR}\xspace}
\def\sdss{\textit{Stripe82}\xspace}
\def\tess{\textit{TESS}\xspace}

\def\ss{\textsc{SuperSmoother}\xspace}
\def\sssp{\textsc{SuperSmootherSP}\xspace}
\def\ls{\textsc{Lomb-Scargle}\xspace}

\def\platforma{\textsc{PlatformA}\xspace}
\def\platformb{\textsc{PlatformB}\xspace}

\def\ssoglobal{\textsc{SSO-Global}\xspace}
\def\ssosmthread{\textsc{SSO-Thread}\xspace}
\def\ssocascade{\textsc{SSO-Cascade}\xspace}

\def\ssspglobal{\textsc{SSSP-Global}\xspace}
\def\ssspcoalesced{\textsc{SSSP-Coalesced}\xspace}
\def\ssspsmthread{\textsc{SSSP-Thread}\xspace}

\def\cpusso{\textsc{SSO-CPU}\xspace}
\def\cpusssp{\textsc{SSSP-CPU}\xspace}

\def\singleobjsmall{\textit{ObjSmall}\xspace}
\def\singleobjlarge{\textit{ObjLarge}\xspace}
\def\singleobjnt{\textit{Syn3k}\xspace}

\journal{Journal of \LaTeX\ Templates}

\biboptions{authoryear}

\begin{document}

\begin{frontmatter}

\title{GPU-Enabled Searches for Periodic Signals of Unknown Shape}


\author[NAU1]{Michael Gowanlock\corref{mycorrespondingauthor}}
\cortext[mycorrespondingauthor]{Corresponding author}
\ead{michael.gowanlock@nau.edu}
\author[ASU]{Nathaniel R. Butler}
\author[NAU2]{David E. Trilling}
\author[NAU2]{Andrew McNeill}

\address[NAU1]{School of Informatics, Computing, and Cyber Systems, Northern Arizona University, Flagstaff, AZ, 86011, USA}
\address[ASU]{School of Earth \& Space Exploration, Arizona State University, Tempe, AZ, 85287, USA}
\address[NAU2]{Department of Astronomy \& Planetary Science, Northern Arizona University, Flagstaff, AZ, 86011, USA}


\begin{abstract}
Recent and future generation observatories will enable the study of variable astronomical phenomena through their time-domain capabilities. High temporal fidelity will allow for unprecedented investigations into the nature of variable objects --- those objects that vary in brightness over time. A major bottleneck in data processing pipelines is constructing light curve solutions for catalogs of variable objects, as it is well-known that period finding algorithms are computationally expensive. Furthermore, there are many period finding algorithms that are often suited for specific science cases. In this paper, we present the first GPU-accelerated Super Smoother algorithm. Super Smoother is general purpose and uses cross-validation to fit line segments to a time series, and as such, is more computationally expensive than other algorithms, such as Lomb-Scargle. Because the algorithm requires making several scans over the input time series for a tested frequency, we also propose a novel generalized-validation variant of Super Smoother that only requires a single scan over the data. We compare the performance of our algorithms to analogous parallel multi-core CPU implementations on three catalogs of data, and show that it is generally advantageous to use the GPU algorithm over the CPU counterparts. Furthermore, we demonstrate that our single-pass variant of Super Smoother is roughly equally as accurate at finding correct period solutions as the original algorithm. Our software supports several features, such as batching the computation to eliminate the possibility of exceeding global memory on the GPU, processing a single object or batches of objects, and we allow for scaling the algorithm across multiple GPUs.
\end{abstract}


\end{frontmatter}


\section{Introduction \& Background}\label{sec:introduction}
Finding the periodic signals of variable objects in astronomical catalogs is a computationally expensive task. New telescope facilities, such as the Rubin Observatory, will have significant time-domain capabilities, where the magnitudes of objects will be recorded with high frequency over long time scales. Consequently, the astronomy community will be largely interested in finding the periods of variable objects from near-future large synoptic surveys, such as the Vera Rubin Legacy Survey of Space and Time~\citep[LSST;][]{2009arXiv0912.0201L}.

The astronomy community has experience with processing large catalogs of light curve data, such as those produced by the Zwicky Transient Facility~\citep[ZTF;][]{ZTF}, Asteroid Terrestrial-impact Last Alert System~\citep[ATLAS;][]{ATLAS}, All Sky Automated Survey for Supernovae~\citep[ASAS-SN;][]{ASAS-SN}, Catalina Real-Time Transient Survey~\citep[CRTS;][]{CRTS}, and the Panoramic Survey Telescope and Rapid Response System~\citep[Pan-STARRS1;][]{Pan-STARRS1}.

We summarize three factors will impact the computational cost of period finding on large astronomical catalogs: $(i)$ there will be a large number of objects to consider; $(ii)$ there will be potentially large frequency spaces, due to the vast range of scientific cases (searches may span hours to years); and, $(iii)$ there will be a potentially large number of data points to consider in each object's light curve. The number of data points in a light curve and the number of frequencies searched directly impact the computational complexity of all period finding algorithms.

There are many period finding algorithms in the literature~\citep[see, for example,][]{1976Ap&SS..39..447L,1978ApJ...224..953S,1982ApJ...263..835S,dworetsky1983period,friedman1984variable,schwarzenberg1989advantage,reimann1994frequency,schwarzenberg1996fast,palmer2009fast,zechmeister2009generalised,2010ApJS..191..247T,mcwilliam2011rr,huijse2012information,2013MNRAS.434.2629G}. These algorithms target different aspects of period finding, such as reducing the computational complexity of the algorithm, parallelizing the algorithm, or improving the probability that an algorithm is able to correctly identify a periodic signal in a time series. One key difference among algorithms is that some are phased-based and require sorting the time series by a searched period, such as \ss~\citep{friedman1984variable}, whereas others such as \ls~\citep{1976Ap&SS..39..447L,1982ApJ...263..835S}, are not phased-based and thus do not require this pre-processing step. 

\begin{figure*}[!t]
\centering
\includegraphics[width=0.48\textwidth]{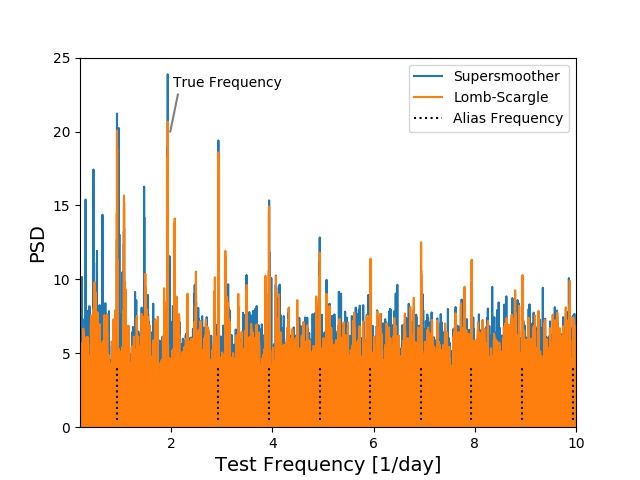}
\includegraphics[width=0.48\textwidth]{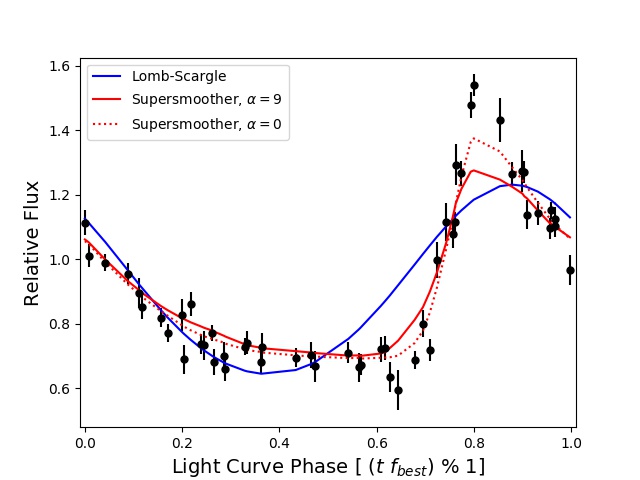}
  \caption{\emph{Left:} The periodogram derived for an RR-Lyrae variable star from the SDSS survey.  There are multiple candidate frequencies because the daily observing period is near the true source frequency (1.9 day$^{-1}$).  \emph{Right:} The identification of the true period is accomplished by fitting models that can account for the source flux variation, which is not simply sinusoidal. The algorithm uses a parameter $\alpha$ which can take a value in the range [0, 10], and we show $\alpha=0$ and $\alpha=9$.}
   \label{fig:period_selection_rr}
   \end{figure*}

\ls has a low computational complexity compared to other algorithms and has been widely utilized in scientific software packages, such as SciPy~\citep{SciPy} and AstroPy~\citep{AstroPy}. In contrast to \ls that fits a sinusoid to the data to find periodic signals, \ss uses cross-validation to fit the data using line segments. Consequently, periodic signals that have greater structure and that are not well-represented by a sinusoid may be better suited to smoothing algorithms, such as \ss. In addition, most ground-based observatories have a 24-hour observing cadence. This cadence can directly produce a periodic artifact in the periodogram, where period finding algorithms detect this 24-hour signature, which may falsely report that the period of an object is likely to be 24-hours (or an alias, such as 12 or 48 hours). An example of this is demonstrated in Figure~\ref{fig:period_selection_rr}, which shows the periodogram of an RR-Lyrae star derived by \ls and \ss. Here, we observe that the power is greater for \ss than \ls at the true frequency (1.9 day$^{-1}$), where the former algorithm has a higher probability of deriving the true period than the latter. Given a large catalog of objects from a ground-based observatory, it may be preferable to employ \ss over \ls despite its greater computational cost.

Similarly to other works that utilize GPUs to improve the performance of data processing pipelines for large-scale surveys~\citep{katz2021gpu,van2021ztf,coughlin2021ztf}, this paper proposes a GPU-accelerated \ss algorithm, and a more computationally efficient variant. Since each searched frequency can be computed in parallel, the GPU is an excellent architecture for carrying out this computation. In summary, this paper makes the following contributions.

\begin{itemize}
\item We propose the first GPU-accelerated \ss algorithm in the literature.
\item To address the high computational complexity of \ss, we propose a single-pass variant of the algorithm. This algorithm uses generalized-validation instead of cross-validation, and makes fewer scans over an object's light curve. We denote this variant as \sssp.
\item To exploit parallel architectures, each frequency must be searched in parallel. This necessitates replicating the dataset for each searched frequency and sorting it by the phase derived from the searched period. This significantly increases the memory footprint of the algorithm which can exceed the GPU's global memory capacity. To address this limitation, we incorporate a batching scheme that splits the search frequencies into several discrete batches, which obviates GPU memory limitations.
\item We optimize the original and single-pass variant for execution on the GPU. Optimizations include exploiting shared-memory and modifying the memory access patterns to exploit coalesced global memory accesses.
\item Our software allows for scaling the computation of a catalog of objects, or a single object, across multiple GPUs.
\item We evaluate our algorithm on two platforms, one with a single GPU, and one with four GPUs, and show performance across two single object datasets and three real-world catalogs, \sdss, \tess, and \linear. These catalogs span a small, medium, and large number of objects, respectively.
\end{itemize}

The paper is organized as follows. Section~\ref{sec:ss_original} describes the original \ss algorithm and our associated GPU implementation, and Section~\ref{sec:ss_variant} presents our single-pass variant of the algorithm. Section~\ref{sec:summary_modes} summarizes the features of our software and supported modes of operation. Section~\ref{sec:exp_eval} experimentally evaluates the algorithm on several datasets and experimental scenarios. Finally, Section~\ref{sec:conclusions} concludes the work and describes future research directions.

\section{GPU Version of Friedman's Super Smoother}\label{sec:ss_original}
\subsection{Overview and Challenges}\label{sec:overview}

The \ss algorithm performs local linear regression to fit line segments to data points. In the astronomical context, the data points are typically defined as time, magnitude, and photometric error. Using several bandwidths, it uses cross-validation to determine the best fitting line segment for a local subset of the data. In its original form, the \ss algorithm requires making nine passes (smooths) over the input dataset for a given frequency, which makes it much more computationally expensive than other period finding algorithms that derive periods in a single pass for a given frequency, such as the single pass Lomb-Scargle Algorithm~\citep{1976Ap&SS..39..447L,1982ApJ...263..835S,press1992numerical,2010ApJS..191..247T}. 

Here, our objective is to search frequencies in parallel using a single GPU thread per frequency, thus exploiting the GPU's massive parallelism. However, a major concern for the parallelization and performance of the \ss algorithm on the GPU is the algorithm's space complexity. The memory footprint scales with frequency and number of data points for the following elements of the algorithm: 

\begin{itemize}
\item \textit{sc}-  Scratch space: $8N_tN_f$ 
\item \textit{smo}-  Output smooth array: $N_tN_f$     
\item \textit{freqarr}-  Bookkeeping of frequency ids for the back-to-back sort: $N_tN_f$
\item \textit{argkeys}-  Keys sorted by values in the back-to-back sort: $N_tN_f$
\item \textit{t1}-  Time after sorting phased time series: $N_tN_f$
\item \textit{t1Asort}-  Time after argsorting t1: $N_tN_f$
\item \textit{yAsort}-  Data after argsorting by t1: $N_tN_f$
\item \textit{wAsort}-  Weights after argsorting by t1: $N_tN_f$
\end{itemize}

Note that the largest contribution to the memory footprint is scratch space. 
The total space complexity of the above components with constant factors is: $15N_tN_f$. With the exception of working memory for the back-to-back sorting operation, the other components of the algorithm require minimal additional space relative to the above factors. 

As an illustrative example, consider an object with $N_t=50$ data points, and a frequency grid with $N_f=10^6$ searched frequencies (typical sizes for $N_t$ and $N_f$). Assuming the data are stored in 64-bit floating point precision (FP64), the space complexity is 5.59 GiB. Modern GPUs have roughly 5--20 GiB of on-card memory, so period searches with a large number of frequencies and/or data points may be intractable due to memory limitations.




\subsection{GPU Implementation Details}\label{sec:ssimplementation} 
As a baseline for comparison, we directly implement Friedman's \ss algorithm~\citep{friedman1984variable} using the GPU. Because the algorithm has to pre-process the data and perform the sorting procedure, the algorithm requires the execution of several kernels. Algorithm~\ref{alg:algssoverview} provides an overview, and we present the main GPU kernel in Listing~\ref{lst:kernelss}. In the pseudocode, parameters that are computed in a given GPU kernel are shown in bold face, whereas inputs to the kernel are shown without bold face. In Algorithm~\ref{alg:algssoverview}, only the compute\_chi0\_tt\_weights and findPeriod functions are computed on the host. All other operations are computed on the GPU as CUDA kernels.

\begin{algorithm*}[!t]
\caption{Pseudocode overview of \ss.}
\label{alg:algssoverview}
\begin{algorithmic}[1]

\begin{footnotesize}
\Procedure{\ss}{$N_t$, $N_f$, $t$, $y$, $e$}
\State $chi0$, $tt$, $w$ $\leftarrow$ compute\_chi0\_tt\_weights($N_f$, $t$, $y$, $e$) \label{algline:chi0ttweights}
\State transferDataHostToGPU($tt$, $y$, $w$) \label{algline:htod}
\State computePhasedTime($N_t$, $N_f$, $f_{min}$, $df$, $tt$, {\boldmath{$t1$}}) \label{algline:periodmodf}
\State initKeyArrays(\textbf{argkeys}, \textbf{freqArr})\label{algline:initarrs}
\State argsort($t1$, freqArr, \textbf{argkeys}) \label{algline:argsort}
\State mapArgsort(argkeys, $t1$, $y$, $w$, \textbf{t1Asort}, \textbf{yAsort}, \textbf{wAsort}) \label{algline:mapargsort}
\State supsmu($N_t$, $N_f$, iper, span, alpha, t1Asort, dataAsort, weightsAsort, sc, \textbf{smo}) \label{algline:supsmu}
\State computePgram($N_t$, $N_f$, $chi0$, $smo$,  dataAsort, weightsAsort, \textbf{pgram}) \label{algline:computepgram}
\State transferDataGPUToHost(pgram)\label{algline:dtoh} 
\State p $\leftarrow$ findPeriod($N_f$, pgram, $\Delta f$)\label{algline:computeperiod}
\State \Return{p}
\EndProcedure

\end{footnotesize}
\end{algorithmic}
\end{algorithm*}

Algorithm~\ref{alg:algssoverview} takes as input the length of the time series, $N_t$, the number of frequencies $N_f$, the time ($t$), magnitude ($y$), and photometric error ($e$). The algorithm returns the period corresponding to the frequency with the highest power. On line~\ref{algline:chi0ttweights}, the algorithm computes the \texttt{chi0} term which will be used to compute the periodogram, an array \texttt{tt} which is the time, $t$, after the minimum time has been subtracted, and a weights array ($w$) computed as $e^{-2}$. Next, on line~\ref{algline:htod}, the algorithm transfers $tt$, $y$, and $w$ to the GPU.  Line~\ref{algline:periodmodf} computes $t1$ as follows: \texttt{t1=(tt~mod~p)/p}. Line~\ref{algline:initarrs} initializes the \texttt{argkeys} and \texttt{freqArr} arrays, where for each frequency, \texttt{argkeys} is initialized with values $0, 1, \ldots,N_t-1$, and \texttt{freqArr} is initialized using the frequency number for $N_t$ elements. Next, line~\ref{algline:argsort} performs an argsort on $t1$. Because we perform the argsort for all frequencies examined at once instead of executing several smaller independent sorts for each frequency, this requires a back-to-back sort. The sort uses $t1$, \texttt{argkeys}, and  \texttt{freqArr} where \texttt{argkeys} is the ordered set of keys after argsorting by $t1$, and \texttt{freqArr} is used to map each index to its respective frequency in \texttt{argkeys}.  The sort is implemented using two calls to the Thrust library's \verb|thrust::stable_sort_by_key()| function~\citep{bell2012thrust}, which is  a state-of-the-art GPU sorting algorithm. Next, the main \ss kernel is called on line~\ref{algline:supsmu}, which is outlined in Listing~\ref{lst:kernelss} below. The kernel takes as input several parameters described above, and outputs \texttt{smo}, an array of smoothed values. Using \texttt{smo}, the periodogram is computed on line~\ref{algline:computepgram}, and is then returned to the host on line~\ref{algline:dtoh}. Finally, the period is computed on line~\ref{algline:computeperiod}, which performs an $\argmax_x(pgram)$ to find the index, $x$, in the periodogram with the maximum power and the period is computed as $p=(f_{min}+x\Delta f)^{-1}$.

We briefly explain how all of the CUDA kernels have been parallelized. The \texttt{computePeriodModF}, \texttt{initKeyArrays}, and \texttt{mapArgsort} kernels all use $N_fN_t$ threads. The \texttt{computePgram} kernel uses $8N_f$ threads. Lastly, as a baseline, our global memory \ss kernel, \texttt{supsmu}, uses $N_f$ threads. 


Listing~\ref{lst:kernelss} presents the main \ss kernel which is directly ported from Friedman's algorithm~\citep{friedman1984variable}. Note that as a baseline we simply use global memory and do not store any information in shared memory. The reason for this is twofold. First, each frequency has a different set of data that it needs to iterate over; therefore, threads cannot share information, thus limiting the potential benefit of data reuse. Second, while we can cache data elements into shared memory for each thread, such as the \texttt{t1Asort}, \texttt{dataAsort}, and \texttt{weightAsort} arrays, this can utilize a significant amount of shared memory. Depending on the size of $N_t$, it is possible that too much shared memory could be requested which would result in a kernel launch failure. 

In Listing~\ref{lst:kernelss}, we show all floating point values being stored in 64-bit precision. However, the source code allows the user to toggle between using FP32 or FP64 using a C macro. For clarity, we did not present the macro in the listing.

In this paper, we assume that all objects are independent of each other, and that they do not share the same time sampling.  Typically, for all sky surveys, which is the primary motivator for this work, there may be a low probability that objects share the same telescope pointings for their entire observational records.  However, in the case where a targeted astronomical campaign were to observe objects in the same field and generate a single time sampling appropriate for all objects, it would be possible to reuse the same folded time intervals for all objects. We do not examine that optimization here, as it breaks the assumption that all objects are independent, and as we will show in Section~\ref{sec:exp_eval}, the folding step of the algorithm is not a bottleneck relative to the other algorithm components.

\begin{lstlisting}[language=C, basicstyle=\footnotesize, floatplacement=tbp, caption=Listing of the \ssoglobal global memory CUDA kernel. \label{lst:kernelss}]
__global__ void SSO-Global(const int N_t, const int N_f, 
const int iper, const double span, const double alpha, 
double * t1Asort, double * dataAsort, 
double * weightsAsort, double * sc, double * smo)
{
    unsigned int tid=threadIdx.x+(blockIdx.x*blockDim.x); (*@\label{lstss:tid}@*)
    if (tid>=numThreads) return; (*@\label{lstss:if}@*)
	
    //offsets into arrays for the thread
    const unsigned int dataOffset=tid*n;
    const unsigned int scOffset=tid*n*8;

    //pointers to time, data, weights 
    //offsets for smo and sc
    double * x=t1_sortby_argkeys+dataOffset;
    double * y=data_sortby_argkeys+dataOffset;
    double * w=weights_sortby_argkeys+dataOffset;
    double * smo_thread=smo+dataOffset;
    double * sc_thread=sc+scOffset;
    int i,j,jper;
    double vsmlsq,sw,sy,a,scale,resmin,tmp,f;

    //spans to be estimated: tweeter, midrange, and woofer
    double spans[] = {0.05,0.2,0.5};
    
    if (x[n-1]<=x[0]) {
      sy=0.0;
      sw=sy;
      for (j=0;j<n;j++) {
        sy=sy+w[j]*y[j];
        sw=sw+w[j];
      }
      a=0.0;
      if (sw>0) a=sy/sw;
      for (j=0;j<n;j++) smo_thread[j] = a;
      return;
    }

    i=n/4-1;
    j=3*(i+1)-1;
    scale=x[j]-x[i];
    vsmlsq=1.e-6*scale*scale;

    jper=iper;
    if (iper==2 && (x[0]<0 || x[n-1]>1)) jper=1;
    if (jper<1 || jper>2) jper=1;
    if (span>0) {
      //Fixed span
      smoothkernel(n,x,y,w,span,jper,vsmlsq,
      smo_thread,sc_thread);      
      return;
    }

    //If we made it here, the span will be 
    //estimated and variable
    for (i=0;i<3;i++) {
      smoothkernel(n,x,y,w,spans[i],jper,vsmlsq,
      sc_thread+2*i*n,sc_thread+6*n);
      smoothkernel(n,x,sc_thread+6*n,w,spans[1],-jper,
      vsmlsq,sc_thread+(2*i+1)*n,sc_thread+7*n);
    }

    for (j=0;j<n;j++) { (*@\label{lstss:loopastart}@*)
      resmin=1.e20;
      for (i=0;i<3;i++) {
        if (sc_thread[j+(2*i+1)*n]<resmin) {
          resmin=sc_thread[j+(2*i+1)*n];
          sc_thread[j+6*n]=spans[i];
        }
      }
      if (alpha>0 && alpha<=10 && 
      resmin<sc_thread[j+5*n] && resmin>0) {
        tmp = resmin/sc_thread[j+5*n];
        if (tmp<1.e-7) tmp=1.e-7;
        sc_thread[j+6*n]+=(spans[2]-sc_thread[j+6*n])*
        pow(tmp,10.0-alpha);
      }
    } (*@\label{lstss:loopaend}@*)

    smoothkernel(n,x,sc_thread+6*n,w,spans[1],-jper,vsmlsq,
    sc_thread+n,sc_thread+7*n);

    for (j=0;j<n;j++) {
      if (sc_thread[j+n]<=spans[0]) sc_thread[j+n]=spans[0];
      if (sc_thread[j+n]>=spans[2]) sc_thread[j+n]=spans[2];
      f=sc_thread[j+n]-spans[1];
      if (f<0) {
        f/=spans[0]-spans[1];
        sc_thread[j+3*n]=(1.0-f)*sc_thread[j+2*n]+f*
        sc_thread[j];
      } else {
        f/=spans[2]-spans[1];
        sc_thread[j+3*n]=(1.0-f)*sc_thread[j+2*n]+
        f*sc_thread[j+4*n];
      }
    }
    smoothkernel(n,x,sc_thread+3*n,w,spans[0],-jper,vsmlsq,
    smo_thread,sc_thread+7*n);
    return;
} //end of GPU kernel
\end{lstlisting}

Now that we have outlined the baseline implementation of the original \ss algorithm, we describe three kernel designs, including the baseline global memory kernel and a shared memory kernel. As we will discuss, the shared memory kernel may fail to launch due to memory limitations. Therefore, we will show that both kernels are needed to make the original \ss algorithm achieve good performance and be robust to the size of the light curves, $N_t$. For brevity, we do not detail the code listings for each kernel here, but we refer the interested reader to our open source code repository for further information.

\subsection{Global Memory Kernel: A Baseline}\label{sec:original_global_baseline}
The global memory kernel, \ssoglobal, was described in Listing~\ref{lst:kernelss}. The kernel only uses global memory and is a baseline for which to compare to our other kernels.

\subsection{Shared Memory Kernel: One Thread Per Frequency}\label{sec:original_smthread}
The original \ss algorithm needs to make several passes over the input dataset and writes intermediate data to a scratch space buffer that is allocated for each frequency, $N_f$. This requires many accesses to global memory. Because the accesses to the scratch space are data-dependent and are not well constrained, it is not possible to achieve good coalesced memory accesses in a global memory kernel. To address this limitation, the shared-memory kernel caches data in shared-memory to reduce the latency of global  memory accesses. Each thread is assigned a single frequency to process. For each thread in a CUDA block, we allocate scratch space for the time, data, and weights in shared memory (where the pointers \texttt{x, y, z} in Listing~\ref{lst:kernelss} will reference shared memory). In addition, we store the three spans in shared-memory to reduce register usage, but we do not store other temporary and constant variables so that we do not increase pressure on the capacity of shared memory.  Since the original \ss algorithm requires making nine passes over the dataset, this optimization eliminates a significant number of (unconstrained) global memory accesses. We denote this CUDA kernel as \ssosmthread.

\subsection{Cascade Mode: Addressing Shared-Memory Limitations}\label{sec:cascade}
We will experimentally show that the best performance is achieved using the \ssosmthread kernel; however, as described above, the \ssosmthread can fail to execute due to insufficient shared memory. Therefore, we propose an execution option that attempts to execute the \ssosmthread kernel and then the \ssoglobal kernel if \ssosmthread fails to execute\footnote{Note that we attempted another kernel design that uses a single block to compute a frequency which could be executed between \ssosmthread and \ssoglobal. However, this kernel performed significantly worse than the \ssosmthread and \ssoglobal kernels. Therefore, we do not present its implementation in this paper.}. This ensures that the \ss algorithm achieves the best performance possible while being robust to the size of the light curve, $N_t$. Additionally, since we check for CUDA kernel launch failures at runtime, the algorithm is also robust to differences in hardware platforms that have different shared memory sizes. This ensures that future generations of GPUs are supported by the software.

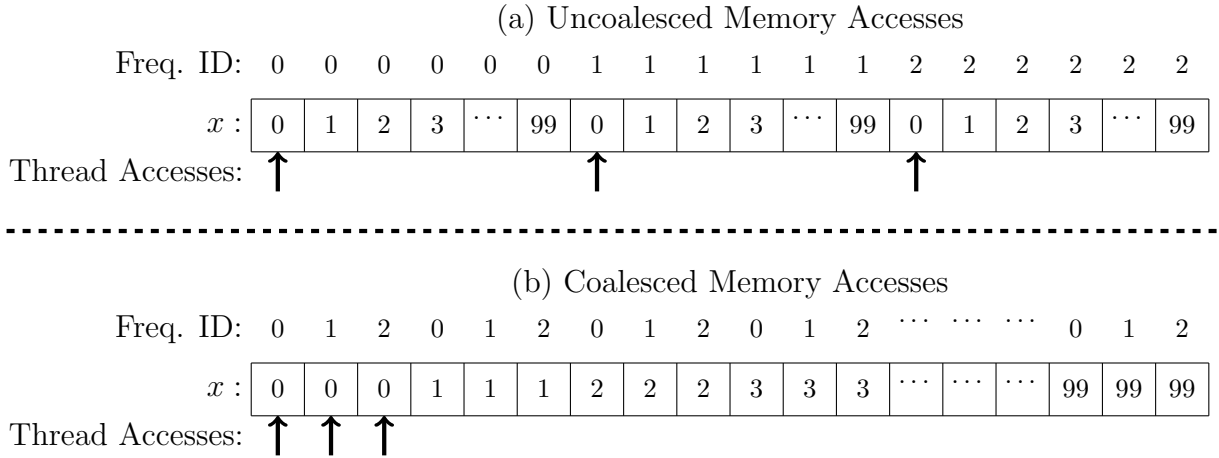
\begin{figure*}[!t]
\centering
  \begin{tikzpicture}[scale=0.7]

\newcommand*{\xMin}{0}%
\newcommand*{\xMax}{18}%
\newcommand*{\yMin}{0}%
\newcommand*{\yMax}{1}%
\def\scale{1.0}

\def\padding{1010101010101010}
\def\keya{\color{red}1111111100000000}
\def\keyb{\color{blue}0000000011111111}


\def\voffseta{0.0}

\node at (9,2.5) {\large{(a) Uncoalesced Memory Accesses}};

\node at (-1.36,1.64) {\large{Freq. ID: }};

\foreach \j\i in {0/0,0/1,0/2,0/3,0/4,0/5,1/6,1/7,1/8,1/9,1/10,1/11,2/12,2/13,2/14,2/15,2/16,2/17} {
        \draw [black] node [below] at (\i+0.5,\yMin+2) {$\j$};
    }

\node at (-0.5,0.5) {\large{$x:$ }};

 \foreach \i in {\xMin,...,\xMax} {
        \draw [black] (\i,\yMin) -- (\i,\yMax); 
    }   
 \foreach \i in {\yMin,...,\yMax} {
    \draw [black] (\xMin,\i) -- (\xMax,\i) node [left] at (\xMin,\i) {};
}

\foreach \j\i in {0/0,1/1,2/2,3/3,$\ldots$/4,99/5,0/6,1/7,2/8,3/9,$\ldots$/10,99/11,0/12,1/13,2/14,3/15,$\ldots$/16,99/17} {
        \draw [black] node [below] at (\i+0.5,\yMin+0.85) {$\j$};
    }

\node at (-2.35,-0.35) {\large{Thread Accesses: }};
\draw [->, ultra thick] (0.5,-0.75) -- (0.5,0);
\draw [->, ultra thick] (6.5,-0.75) -- (6.5,0);
\draw [->, ultra thick] (12.5,-0.75) -- (12.5,0);

\def\voffsetb{-5}

\draw [dashed, ultra thick] (-4.9,\voffsetb+3.5) -- (\xMax+0.2,\voffsetb+3.5);

\node at (9,2.5+\voffsetb) {\large{(b) Coalesced Memory Accesses}};

\node at (-1.36,1.64+\voffsetb) {\large{Freq. ID: }};

\foreach \j\i in {0/0,1/1,2/2,0/3,1/4,2/5,0/6,1/7,2/8,0/9,1/10,2/11,$\ldots$/12,$\ldots$/13,$\ldots$/14,0/15,1/16,2/17} {
        \draw [black] node [below] at (\i+0.5,\yMin+\voffsetb+2) {$\j$};
    }

\node at (-0.5,0.5+\voffsetb) {\large{$x:$ }};

 \foreach \i in {\xMin,...,\xMax} {
        \draw [black] (\i,\yMin+\voffsetb) -- (\i,\yMax+\voffsetb); 
    }   
 \foreach \i in {\yMin,...,\yMax} {
    \draw [black] (\xMin,\i+\voffsetb) -- (\xMax,\i+\voffsetb) node [left] at (\xMin,\i+\voffsetb) {};
}

\foreach \j\i in {0/0,0/1,0/2,1/3,1/4,1/5,2/6,2/7,2/8,3/9,3/10,3/11,$\ldots$/12,$\ldots$/13,$\ldots$/14,99/15,99/16,99/17} {
        \draw [black] node [below] at (\i+0.5,\yMin+\voffsetb+0.85) {$\j$};
    }

\node at (-2.35,-0.35+\voffsetb) {\large{Thread Accesses: }};
\draw [->, ultra thick] (0.5,-0.75+\voffsetb) -- (0.5,0+\voffsetb);
\draw [->, ultra thick] (1.5,-0.75+\voffsetb) -- (1.5,0+\voffsetb);
\draw [->, ultra thick] (2.5,-0.75+\voffsetb) -- (2.5,0+\voffsetb);

\end{tikzpicture}

  \caption{Example of transforming memory accesses to the time, magnitude, and weights arrays with $N_f=3$ and $N_t=100$. For illustrative purposes, we show a search using only $N_f=3$ frequencies.  (a) The data layout and memory access patterns where all data elements for a given frequency are stored contiguously. This memory access pattern leads to uncoalesced memory accesses to global memory. (b) Transforming the memory access pattern in (a) by storing the same data element IDs contiguously across frequencies to achieve coalesced memory accesses.}
   \label{fig:coalesced_tikz}
\end{figure*}

\section{A Single-Pass Generalized Validation Variant of SuperSmoother}\label{sec:ss_variant}   
As described in Section~\ref{sec:ssimplementation}, the original \ss algorithm requires a significant amount of memory for scratch space. To address this problem, we propose a variant of the \ss algorithm. In contrast to using a cross-validation approach as in the original algorithm, our algorithm uses a \emph{generalized validation} approach which only requires a single call to the smooth function and eliminates the use of the scratch space required of the original algorithm. While this approach yields a slightly worse fit to the data compared to the original \ss algorithm, it requires less computation and is still effective for period finding. We will demonstrate that our single-pass algorithm achieves roughly equivalent period finding accuracy as the original algorithm.

Similarly to the original algorithm, we propose several GPU kernel designs. All of the other steps in the algorithm are equivalent, with the exception of the kernel (line~\ref{algline:supsmu} in Algorithm~\ref{alg:algssoverview}). We outline the kernels as follows.

\subsection{Single-pass Analogs of the Original Kernels}\label{sec:single_pass_analogs}
The proposed single-pass kernels are direct analogs of the original kernels outlined in Sections~\ref{sec:original_global_baseline}~and~\ref{sec:original_smthread}. In particular, the \ssspglobal kernel is a baseline that does not utilize any shared memory. The \ssspsmthread kernel assigns a single thread to parallelize each frequency. Each of these kernels have the same characteristics as the kernels in the original algorithms with the exception that only a single smooth operation is required for our generalized validation approach.

\subsection{Global Memory Kernel: Exploiting Coalesced Memory Accesses}\label{sec:single_pass_coalesced}
The original \ss algorithm requires nine calls to the smooth function. A drawback of the original algorithm's \ssoglobal kernel is that it requires scratch space for each frequency that is accessed in a non-deterministic manner. It is well-known that to achieve good performance on the GPU, global memory accesses should be coalesced where possible~\citep{bell2012thrust}. Otherwise, multiple transactions are required to retrieve data from global memory, which does not effectively utilize global memory bandwidth. Incorporating memory access patterns that guarantee coalesced memory accesses into the original \ssoglobal kernel is not possible due to the abovementioned non-deterministic memory access patterns. In contrast, our single-pass variant has deterministic memory access patterns to the time, magnitude, and weights data (\texttt{x\rm{,} y\rm{,} \rm{and} w} in Listing~\ref{lst:kernelss}). We exploit these regularized memory access patterns to create a global memory kernel that has coalesced memory accesses.

Figure~\ref{fig:coalesced_tikz} illustrates transforming the non-coalesced memory accesses patterns into coalesced memory accesses, where $N_f=3$ and $N_t=100$.  Figure~\ref{fig:coalesced_tikz}(a) shows accessing an array (here, $x$ refers to the time, magnitude, or weights arrays), where  elements corresponding to the same frequency are stored contiguously. Therefore when the $N_f=3$ threads access the data, the memory accesses occur with a stride of size $N_t$. This requires three transactions to global memory to occur. In contrast, Figure~\ref{fig:coalesced_tikz}(b) shows accessing the same array, where data elements with the same ID are stored contiguously across the frequencies. Consequently, the $N_f=3$ threads access the data contiguously, and this only requires a single coalesced load from global memory. This reduces the number of global memory loads that are required of the algorithm, thereby effectively utilizing global memory bandwidth.

We refer to the global memory kernel with coalesced memory accesses to the time, magnitude, and weights arrays as \ssspcoalesced. Note that since we modify the data layout, the \texttt{mapArgsort} and \texttt{computePgram} kernels in Algorithm~\ref{alg:algssoverview} on lines~\ref{algline:mapargsort}~and~\ref{algline:computepgram}  need to be modified to use this transformation.

\section{Optimizations and Summary of Modes}\label{sec:summary_modes}
In this section, we outline optimizations common to both the original and single-pass algorithms and summarize the functionality of our software.

\subsection{Batching Scheme}\label{sec:batching}
As described in Section~\ref{sec:overview}, the \ss algorithm requires a significant amount of global memory; therefore, it may be intractable to perform a period search across all frequencies without exceeding global memory capacity. Since the frequencies can be searched independently, we divide the $N_f$ frequencies into $N_b$ batches. We compute $N_b$ by estimating the total global memory footprint, $r$, divided by the global memory size of the device, $s$ (in GiB). The software automatically detects the global memory size of the GPU such that it does not require any user intervention. Because $r$ is an estimate of the total global memory footprint, and because we do not have direct control over when memory will be freed by the system, we underestimate the global memory capacity by a coefficient $\beta$ where $\beta<1$ to avoid exceeding global memory capacity. The number of batches is given as follows:

\begin{equation}
N_b=r\cdot(\beta s)^{-1}.\label{eqn:number_of_batches}
\end{equation}

In this paper, we set $\beta=0.75$\footnote{Note that this default value is useful for a dedicated GPU where memory is not used for other purposes. In cases where the GPU is not dedicated, such as in a laptop that is also rendering a user interface, then it may be preferable to decrease the value of $\beta$. This parameter can be modified in the parameters file in the source code.}. Since the original \ss algorithm requires scratch space and the single-pass variant does not, we require two different values for $r$ for a given level of floating point precision. We denote $r_{orig}$  and $r_{sp}$ as the estimate of the original \ss and the single-pass algorithm's memory footprints, respectively. We estimate these terms in GiB for both FP32 and FP64 precision floating point values as follows.

\begin{itemize}
\item $r_{orig}^{FP32}=1024^{-3}\cdot[(4N_f)+(12N_t)+(84N_fN_t)]$
\item $r_{orig}^{FP64}=1024^{-3}\cdot[(8N_f)+(24N_t)+(160N_fN_t)]$
\item $r_{sp}^{FP32}=1024^{-3}\cdot[(4N_f)+(12N_t)+(52N_fN_t)]$
\item $r_{sp}^{FP64}=1024^{-3}\cdot[(8N_f)+(24N_t)+(96N_fN_t)]$
\end{itemize}





From the estimates of the FP64 memory footprints, the original \ss algorithm requires a factor $1.67$ more memory on the $N_fN_t$ terms than the single pass algorithm due to additional scratch space. This increases the number of batches and kernel invocations that need to be executed. To accommodate the batching scheme, Algorithm~\ref{alg:algssoverview} is modified by including a loop around lines~\ref{algline:periodmodf}--\ref{algline:computepgram} that iterates over the batches from from $1, 2, \ldots, N_b$.

\subsection{Multi-GPU Support} 
We add multi-GPU functionality to the algorithm. This allows us to distribute batches to GPUs, where $N_{GPU}$ denotes the number of employed GPUs. When we use multi-GPU mode, to obtain low load imbalance at the end of the computation, we ensure that the number of batches evenly divides the number of GPUs. Therefore, when we estimate the number of batches when using multi-GPU mode, we use $N_b^\prime$ batches, where $N_b^\prime=\lceil\frac{N_b}{N_{GPU}}\rceil\cdot N_{GPU}$. Therefore, each GPU will compute $N_b^\prime/N_{GPU}$ batches, where $N_b^\prime~\mathrm{mod}~N_{GPU}=0$. To enable the GPUs to execute concurrently, we use $N_{GPU}$ threads on the host, which are parallelized using OpenMP.

There may be the case where GPUs in a multi-GPU system have differing capabilities. In batch mode, each GPU is assigned a different object. In this case, the code dynamically assigns objects to the GPUs. This is because objects can have differing time series lengths, $N_t$, which directly impact execution time. Thus, even when using the same GPUs in a multi-GPU system, the GPUs are unlikely to process the exact same number of objects. Similarly, if using GPUs with varying capabilities, this mechanism will allow the GPUs to dynamically process the objects, where the slower GPU(s) will process fewer objects on average compared to the faster GPU(s). To do this, a minor modification to the code would be needed to accommodate the global memory capacity of each GPU such that the number of batches can be calculated (Equation~\ref{eqn:number_of_batches}). When computing the periodogram of a single object, the code could be modified to dynamically assign batches of work to each GPU where the number of batches (Equation~\ref{eqn:number_of_batches}) is computed using the GPU with the smallest global memory capacity.

\subsection{Summary of Modes and Options}
We summarize the modes and options provided by our software as follows. 

\begin{itemize}
\item The software consists of GPU implementations of Friedman's \ss algorithm and our single-pass variant.
\item The software supports both FP32 and FP64 floating point arithmetic modes. This allows the user to select an appropriate level of precision for their scientific investigation. GPUs designed for scientific computing have a significant amount of hardware dedicated to FP64 whereas consumer grade GPUs have minimal hardware dedicated to FP64; therefore, a user may also wish to select the precision level based on available hardware. 
\item The software also supports processing astronomical catalogs containing different objects. In this mode, we assume the same frequency grid for all objects, as we expect catalogs to target different science cases (e.g., asteroids compared to RR-Lyrae)
\item As described in Sections~\ref{sec:overview}~and~\ref{sec:batching}, the \ss algorithm requires a significant amount of memory to enable parallelization across frequencies. We provide a batching scheme such that the GPU does not exceed global memory capacity. We automatically detect the GPU's global memory capacity, and estimate the algorithm's memory footprint to determine the number of batches that are required to process an object. This ensures that platform-specific details do not need to be entered by the user.
\item Our software supports multi-GPU systems which are becoming increasingly ubiquitous in astronomical data processing pipelines.
\end{itemize}

\section{Experimental Evaluation}\label{sec:exp_eval}

\subsection{Experimental Methodology}\label{sec:exp_method}
All host code is written in C/C++ and all GPU code is written in CUDA. The code is compiled using the O3 optimization flag, and unless otherwise noted, all reported response times are averaged over 3 time trials. \ss takes as input a value of $\alpha$, and in this paper, for period finding purposes, we set $\alpha=9.0$. The parameter is used to tune how sensitive the algorithm is to outlier data points, and we will discuss the selection of this parameter in Section~\ref{sec:alpha_parameter}.

\begin{table*}[!t]
\setlength\tabcolsep{3pt}
\centering
\footnotesize
\caption{Details of the platforms used in the experimental evaluation. The number of CUDA cores and global memory size are shown for each GPU in \platformb. FP refers to the floating point precision capabilities of the GPUs.}\label{tab:platforms}
\begin{tabularx}{\textwidth}{|l|Ylll|Yll|} 
\hline
& \multicolumn{4}{c|}{\textbf{CPU}} &  \multicolumn{3}{c|}{\textbf{GPU}}\\
 Platform& Model & Cores & Clock & Memory &Model & Cores & Memory\\
\hline
\platforma & $2 \times$E5-2620 v4& $2 \times8=16$& 2.1 GHz& 128 GiB& Quadro GP100 & 3584 & 16 GiB\\
\platformb & W-2295& 18& 3.0 GHz& 256 GiB& $4\times$Quadro RTX 5000 & 3072 & 16 GiB\\
\hline
\end{tabularx}
\end{table*}

Table~\ref{tab:platforms} outlines the platforms used in our experimental evaluation. \platforma contains Intel Xeon CPUs with 16 total physical cores and an Nvidia Quadro GP100 GPU (Pascal generation). \platformb contains an Intel Xeon CPU with 18 total physical cores equipped with 4 Nvidia Quadro RTX 5000 GPUs (Turing generation). The Quadro GP100 in \platforma supports both FP32 and FP64 floating point precision. Unless otherwise noted,  when we use this platform, we execute the algorithm with FP64. In contrast, the Quadro RTX 5000 GPUs in \platformb only contain 1/32 the hardware for FP64 as FP32\footnote{\url{https://www.nvidia.com/content/dam/en-zz/Solutions/design-visualization/technologies/turing-architecture/NVIDIA-Turing-Architecture-Whitepaper.pdf}}. The small hardware allocation for FP64 allows the GPU to be able to execute programs that contain FP64 instructions, but it is not designed to execute programs that require a significant amount of FP64 computation.

Due to the large range of GPU kernels described in Sections~\ref{sec:ss_original}~and~\ref{sec:ss_variant}, we summarize the GPU kernels and CPU implementations below.

The kernels associated with the original \ss algorithm on the GPU are as follows:

\noindent\textbullet \ssoglobal -- Global memory baseline (Section~\ref{sec:original_global_baseline}).

\noindent\textbullet \ssosmthread -- Assigns one thread per frequency and uses shared memory (Section~\ref{sec:original_smthread}).

\noindent\textbullet \ssocascade -- Attempts to execute \ssosmthread and if it has insufficient shared memory, executes \ssoglobal (Section~\ref{sec:cascade}).

The kernels associated with the \sssp algorithm on the GPU are as follows:

\noindent\textbullet \ssspglobal -- Global memory baseline (Section~\ref{sec:single_pass_analogs}).

\noindent\textbullet \ssspsmthread -- Assigns one thread per frequency and uses shared memory  (Section~\ref{sec:single_pass_analogs}).

\noindent\textbullet \ssspcoalesced -- Global memory kernel that exploits coalesced memory accesses  (Section~\ref{sec:single_pass_coalesced}).

The parallel CPU implementations are denoted as follows:

\noindent\textbullet \cpusso -- \ss algorithm configured with the same number of threads as physical CPU cores on a platform.

\noindent\textbullet \cpusssp -- \sssp algorithm configured the same as the above.

\subsection{Datasets}\label{sec:datasets}
We outline the datasets used in our experimental evaluation as follows. To evaluate executing our algorithms on a single light curve, we employ a dataset, \singleobjsmall, which is a short light curve of an RR-Lyrae star having $N_t=53$ observations.  We employ a synthetic light curve, denoted as \singleobjlarge, containing $N_t=3,554$ observations corresponding to a large light curve of a synthetic asteroid. The observational record of this object has a cadence similar to that delivered by ZTF~\citep{ZTF}, where we define cadence to be the semi-regular observing pattern. Here, our cadence is based on the ZTF public survey, where data are collected around half of the nights. While we reproduce the semi-regular observing pattern, it does not impact the performance of the algorithms, so it is not important for the purposes of this paper. Lastly, to demonstrate how the algorithms perform as a function of $N_t$, we create a dataset denoted as \singleobjnt, which is the \singleobjlarge dataset that has been partitioned into several lightcurves having $N_t \in [25, 3000]$.

We employ three real-world catalogs of data for demonstrating the performance of the algorithm for batch processing. All of the objects that we examine may be periodic in their brightness.   We use a subsample of 136 RR-Lyrae stars from \citet{2010ApJ...708..717S} that have known period solutions, denoted as \sdss. We employ a dataset of 37,965 asteroid light curves from TESS cycle 1, denoted as \tess~\citep{McNeillInPrep}. In this dataset, asteroid magnitudes vary due to rotation. Finally, we use the \linear dataset, which contains 94,252 sidereal sources, where a subset are variable stars. A typical scenario for finding variable objects in the \linear dataset would be to derive the light curves of these objects and select those that have a sufficiently high normalized power. Afterwards, scanning~\citep{2019Natur.571..528B} and classifying the objects as being periodic or non-periodic sources~\citep{van2021ztf} is needed.  These three catalogs represent small, medium and large sizes, each of which have different light curve length distributions, which will impact the execution time of the algorithms.

\subsection{Selection of The Number of Frequencies}\label{sec:eval_num_freq}
In the experiments that follow, we examine the performance of our algorithms, and this includes observing how performance varies as a function of $N_f$ and $N_t$. It is well-known that in practice, when selecting $N_f$, it is important to select a value that will not miss the peaks in the periodogram~\citep{2018ApJS..236...16V}. Consequently, when we perform our experimental evaluation, we select values of $N_f$ that at least bracket a value of this parameter. In what follows, we outline our procedure for selecting $N_f$ when processing batches/catalogs of objects, which is the main objective of this paper.

We use a similar frequency grid prescription as~\citet{2011ApJ...733...10R}. Let a dataset of objects be denoted as $O$, where $|O|$ is the number of objects in the dataset. Let each object in a given dataset be denoted as $o_i$, where $i=1, 2, \ldots, |O|$. Each object, $o_i\in O$, has a temporal extent denoted as $T_i=T_i^{end}-T_i^{start}$, where $T_i^{start}$ and $T_i^{end}$ refer to the time of the first and last observations of object $o_i$, respectively. We then compute the maximum temporal extent, $T^{max}$, in the dataset/catalog as $T^{max}=\max(T_i)$. We compute the frequency grid spacing as $\Delta f=0.1({T^{max}}^{-1})$. We then compute $N_f$ as follows: $N_f=f_{max}-f_{min}({\Delta f}^{-1})$. Table~\ref{tab:reasonable_nf} shows reasonable values for $\Delta f$ and $N_f$ using the frequency range, $[f_{min}, f_{max})$, where the frequency range is based on the expected periodic signal range for a given dataset which is driven by the properties of the objects. Note that the properties of the \tess dataset are such that we are able to detect light curve periods on the order of tens of days~\citep{2019ApJS..245...29M}, so the applicable period search range is small.

\begin{table}[!t]
\centering
\begin{footnotesize}
\caption{Practical values of $N_f$ for the catalogs that will be processed using batch mode. Frequency ranges are selected based on scientific objective, and in the case of \tess cycle~1,  the survey duration limits the light curve period range.} 
\label{tab:reasonable_nf}
\begin{tabularx}{\columnwidth}{|X|r|r|r|} \hline
Dataset&$[f_{min}, f_{max})$ day$^{-1}$& $\Delta f$ & $N_f$\\\hline
\sdss  &$[0.1, 10.0)$&$3.0\times10^{-5}$&330,000\\
\tess  &$[0.1, 10.0)$&$3.7\times10^{-4}$&26,757\\
\linear&$[0.01, 10.0)$&$4.5\times10^{-5}$&222,000\\
\hline
\end{tabularx}
\end{footnotesize}
\end{table}




\begin{figure*}[!t]
\centering
  \includegraphics[width=1.0\textwidth]{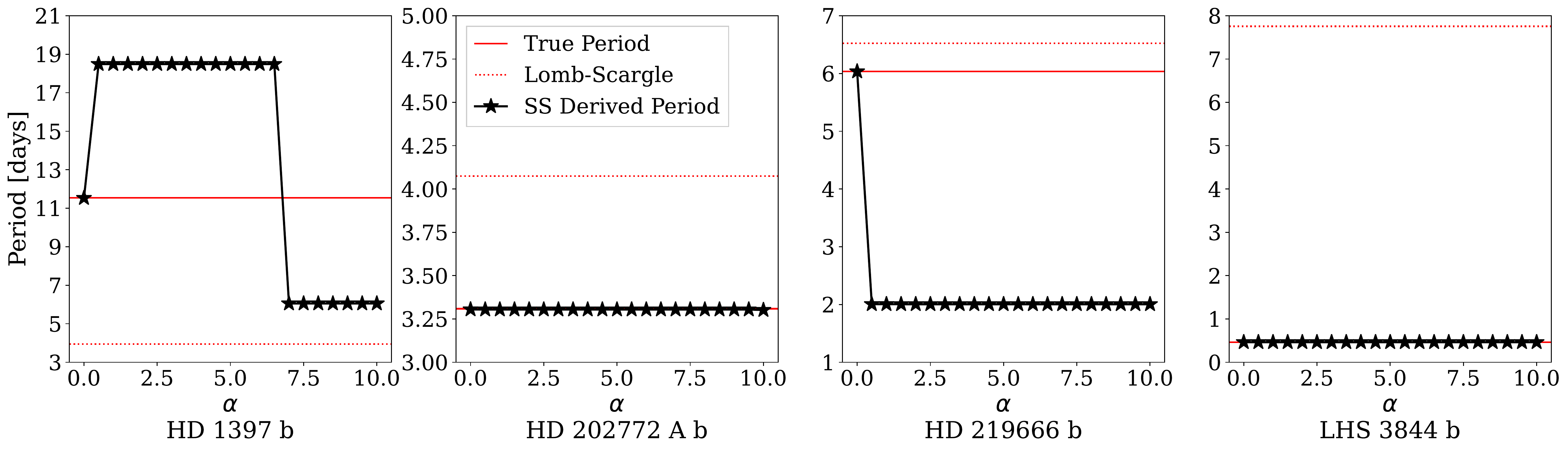}
  \caption{The \ss derived orbital period as a function of the $\alpha$ parameter for the following four exoplanets: HD~1397~b, HD~202772~A~b, HD~219666~b, and LHS~3844~b. Values for the true periods (red solid horizontal line) are obtained from \url{http://exoplanet.eu/}. The dotted red horizontal line shows the periods derived by \ls, which yields incorrect periods in all four cases. The periods were derived using a uniform frequency grid having $N_f=10^6$, and frequencies were searched in the range [0.05, 10.0) day$^{-1}$. Algorithms executed with FP64. }
   \label{fig:TESS-exoplanet}

\end{figure*}

\subsection{Selection of the $\alpha$ Parameter}\label{sec:alpha_parameter}

The $\alpha$ parameter in \ss is used to penalize outliers in the fitting procedure, where  $\alpha=0$ and $\alpha=10$ indicate low and high penalizations, respectively. The benefit of a high value of $\alpha$ is that it will not overfit outliers in the time series, and since astronomical data may have substantial error, these outliers can yield incorrect periods. Figure~\ref{fig:period_selection_rr} demonstrated using two values of $\alpha$. We find that in the case of asteroids and RR-Lyrae, $\alpha=9$  yields good periods, and we use this value when processing catalogs of data with \ss.

We show the case where a low value of $\alpha$ is beneficial over a high value. Transiting exoplanet searches generally use the box least squares (BLS) approach, in which a box filter is scanned over the data to detect transits~\citep{2002A&A...391..369K}. Here, we use \ss to derive the period of an exoplanet transit. Figure~\ref{fig:TESS-exoplanet} shows the orbital periods of four exoplanets derived by \ss as a function of $\alpha$\footnote{These datasets were not described in Section~\ref{sec:datasets}, as we do not use them to evaluate the performance of the algorithm.}. The true period is shown by the horizontal red solid line. From the figure, we observe that selecting  $\alpha=0$ is able to correctly derive all of the orbital periods. This is because in the time series of a transiting exoplanet, the transiting data points are outliers, and we do not want to underfit those data points, otherwise the periodic signal will not be found by the algorithm. For comparison purposes, we also show that \ls is unable to derive any of the correct orbital periods. This is unsurprising, as a pure  sinusoidal fit is unsuitable for fitting the asymmetric steep profile of these light curves; as mentioned above, these light curves are typically fit using BLS and not \ls.

\subsection{Comparison of Frequency Histograms: Lomb-Scargle vs. SuperSmoother}\label{sec:linear_accuracy_SS_LS}
\begin{figure}[!t]
\centering
  \includegraphics[width=1.0\columnwidth]{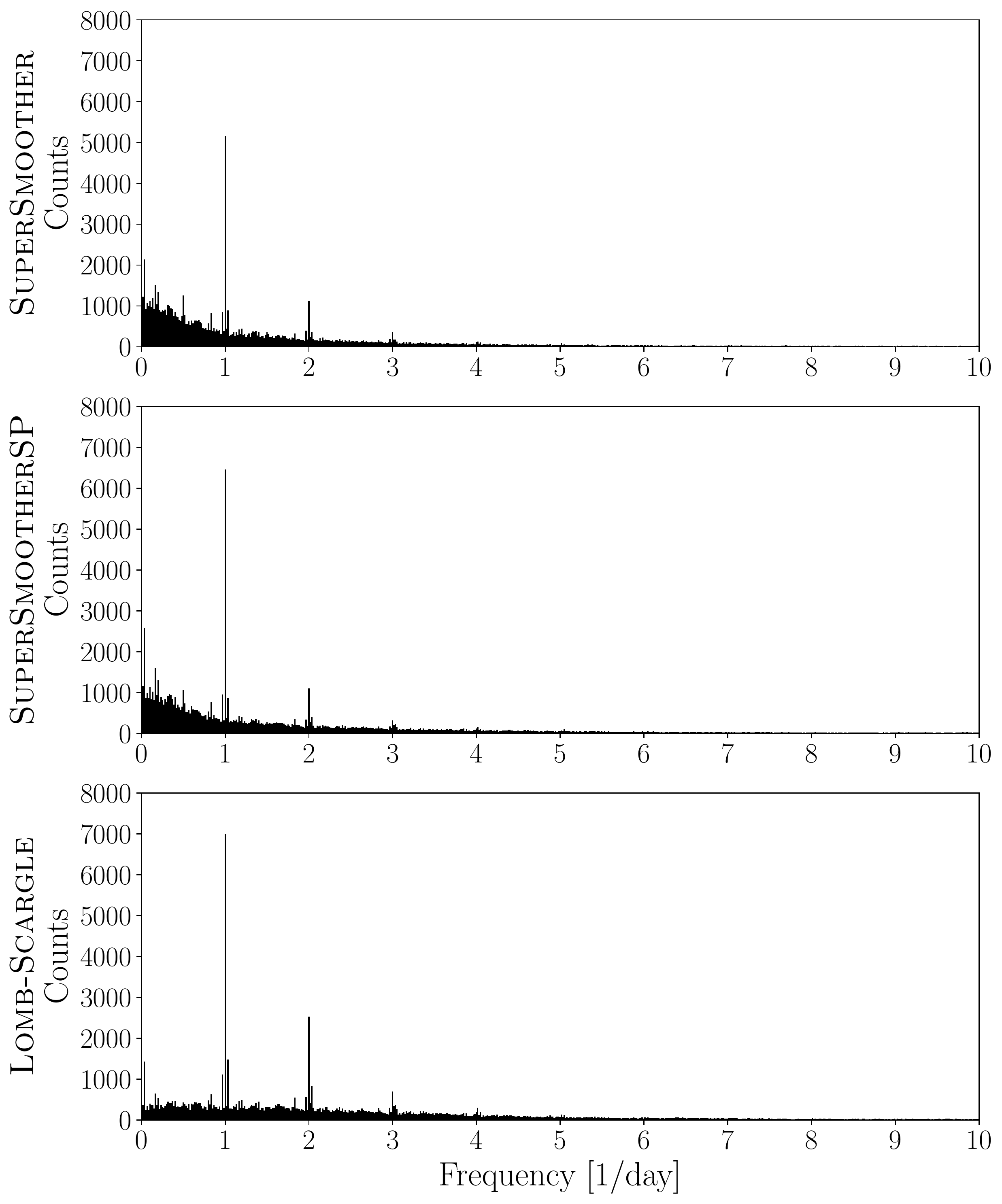}
  \caption{Histograms of the derived frequencies from top to bottom panels using \ss, \sssp, and \ls on the \linear dataset. The periods were derived using a uniform frequency grid having $N_f=222,000$ and frequencies were searched in the range [0.01, 10.0) day$^{-1}$. Algorithms executed with FP32.}
   \label{fig:linearfreqhistogram}
\end{figure}

As described in Section~\ref{sec:introduction}, one benefit of using \ss over \ls is that the algorithm is able to differentiate between periods that are aliases of the 24-hour ground-based telescope observing schedule. Consequently, given an input catalog of objects, we would expect that \ls would have a larger number of period solutions that are aliases of 24-hours. To demonstrate this, Figure~\ref{fig:linearfreqhistogram} plots the derived frequency histograms for \ss and \ls on the \linear dataset. We find that \ss and \sssp have fewer derived periods at frequencies that are multiples of days$^{-1}$ compared to \ls. Since 24-hour aliases will be present for any period finding method, a typical approach is to remove the 24-hour aliased solutions~\citep{coughlin2021ztf}, and instead select a solution that has a lower power. Another application would be to use multiple period finding algorithms (e.g., \ss, \ls, and possibly others) on an input time series to increase confidence in derived period results. This may be another avenue of discerning between 24-hour solutions.

\subsection{Single-pass Variant Accuracy}\label{sec:singlepass_accuracy_SS_LS}
We use the \sdss dataset of RR-Lyrae to examine the accuracy of \ss compared to \sssp, and we also show the accuracy using \ls for comparison purposes. The periods for these RR-Lyrae were derived using light curve templates~\citep{2010ApJ...708..717S}.

\begin{figure}[!t]
\centering
  \includegraphics[width=1.0\columnwidth]{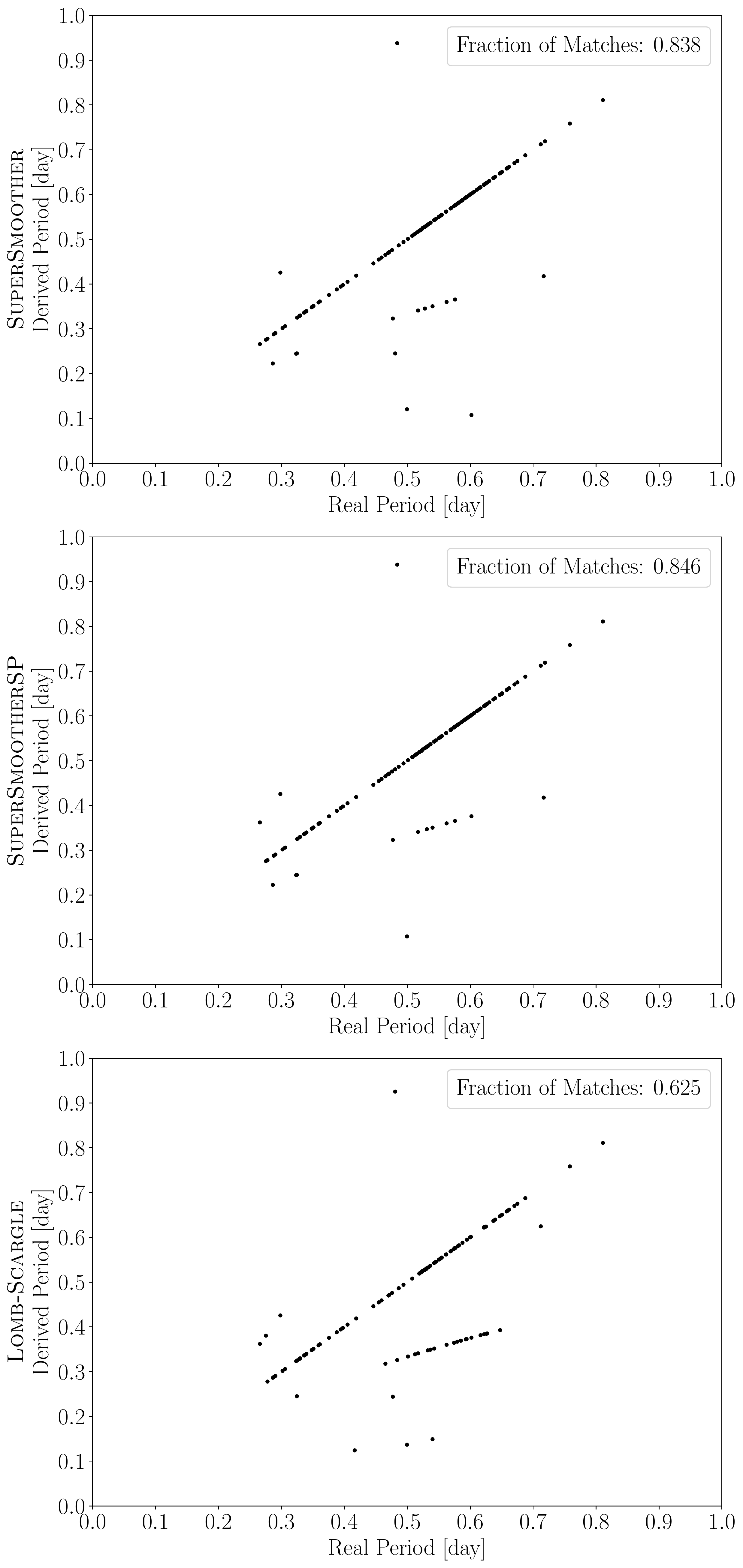}
  \caption{Scatter plots of the derived periods compared to the real periods on the \sdss dataset containing RR-Lyrae stars. Panels from top to bottom are as follows: \ss, \sssp, and \ls. The periods were derived using $N_f=330,000$ in the frequency range [0.1, 10.0) day$^{-1}$. Algorithms executed with FP64.}
   \label{fig:sdss_stripe82_accuracy_three_algs}
\end{figure}

Figure~\ref{fig:sdss_stripe82_accuracy_three_algs} plots the derived periods for \ss (top), \sssp (middle), and \ls (bottom) compared to the real period for the RR-Lyrae in the \sdss dataset. To compute the derived period for a given object, we simply select the frequency having the greatest power value in the periodogram, and do not manually inspect any of the light curves or periodograms.
A perfect match between the derived and real solutions would be indicated by a diagonal line. We compute the fraction of matches between the real and derived solutions, where a match is defined as being within 1\% of the real period. We find that 83.8\% of objects match using \ss and 84.6\% match using \sssp, indicating that the faster single-pass variant does not decrease the ability of the algorithm to find the correct period. We find that \ls finds 62.5\% of the correct periods, demonstrating that \ss and its single-pass variant are more effective at accurately deriving periods on this dataset. This motivates the use of \sssp over \ss, as it is highly effective at deriving periods and has a much lower computational complexity.

\subsection{Scalability of CPU Implementations}\label{sec:eval_cpu_scalability}
\begin{figure}[!t]
\centering
  \includegraphics[width=1\columnwidth]{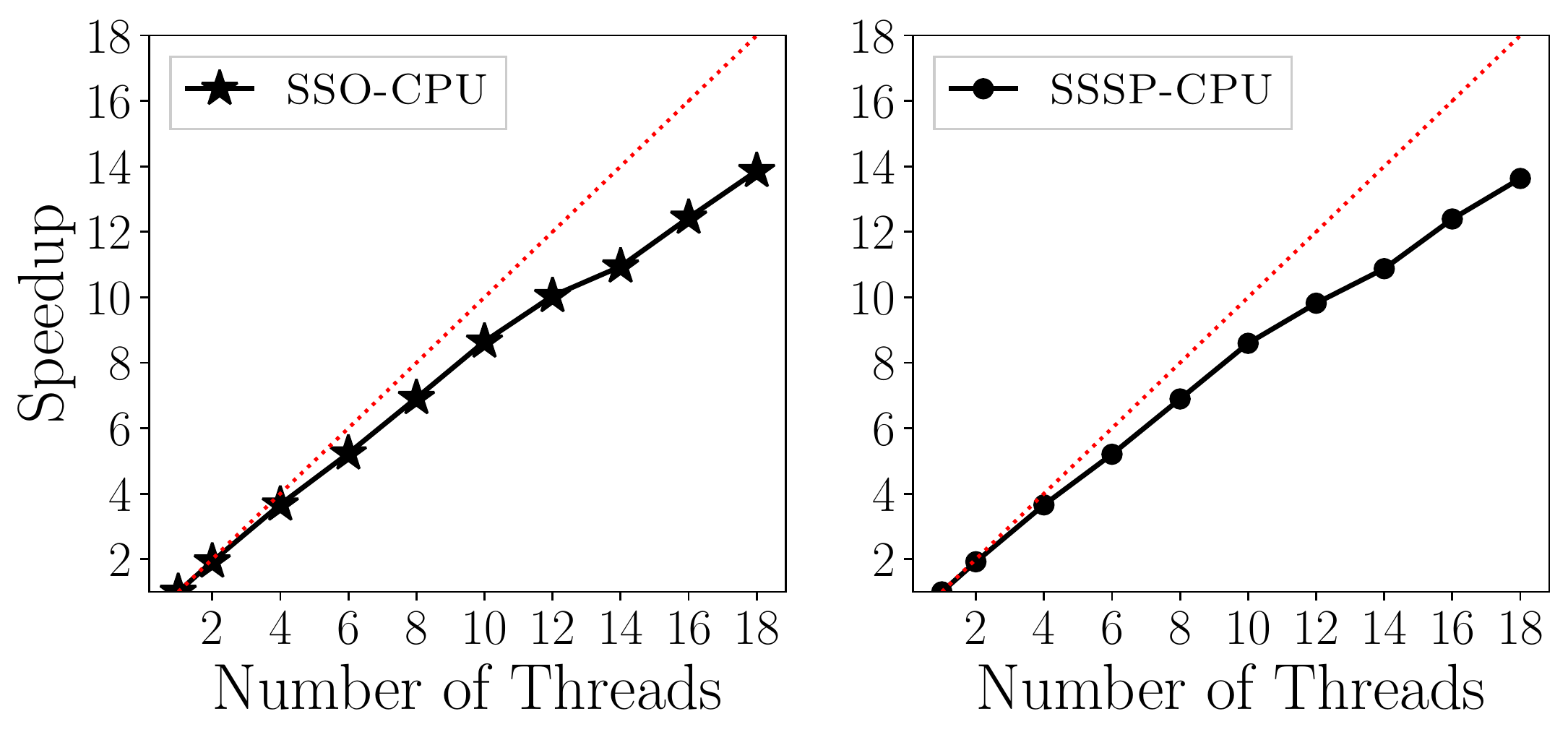}
  \caption{Scalability of the CPU implementations. Speedup as a function of the number of threads on the \singleobjlarge dataset. Frequencies are searched in the range [0.1, 10.0) day$^{-1}$ using $N_f=5\times10^5$. \emph{Left:} \cpusso, and \emph{Right:} \cpusssp. In each plot, a perfect speedup is indicated by the dashed line. Experiments performed on \platformb with FP64.}
   \label{fig:cpu_scalability}
\end{figure}

Before comparing the performance of the CPU and GPU implementations, we assess the scalability of the CPU implementations. In this experiment, we carry out the search on a single object, where the frequencies to be searched  are partitioned between the CPU threads. Figure~\ref{fig:cpu_scalability} plots the speedup compared to the number of threads on the \singleobjlarge dataset. We find that \ss (left panel) and \sssp (right panel) achieve excellent scalability on the CPU yielding a speedup on 18 CPU cores of 13.86$\times$ and 13.63$\times$, respectively. 

Comparing the response times used to generate Figure~\ref{fig:cpu_scalability}, we find that on 18 cores, \ss executes in 19.89~s, whereas \sssp executes in 17.12~s. Thus, the single-pass variant only provides a minor performance advantage on the CPU. Because the \ss algorithm makes several passes over the sorted time series for a given frequency, the time series can be read once from main memory and then stored in cache, and subsequent scans over the time series exploit high data reuse in the cache. In contrast, the single-pass variant does not make multiple scans over the time series, and so it is unable to exploit data reuse in the cache to the same degree as the original algorithm. This explains why the performance is similar between \ss and \sssp on the CPU despite the lower complexity of the latter algorithm.

\subsection{Performance of Cascade Mode}
\begin{figure}[!t]
\centering
  \includegraphics[width=1.0\columnwidth]{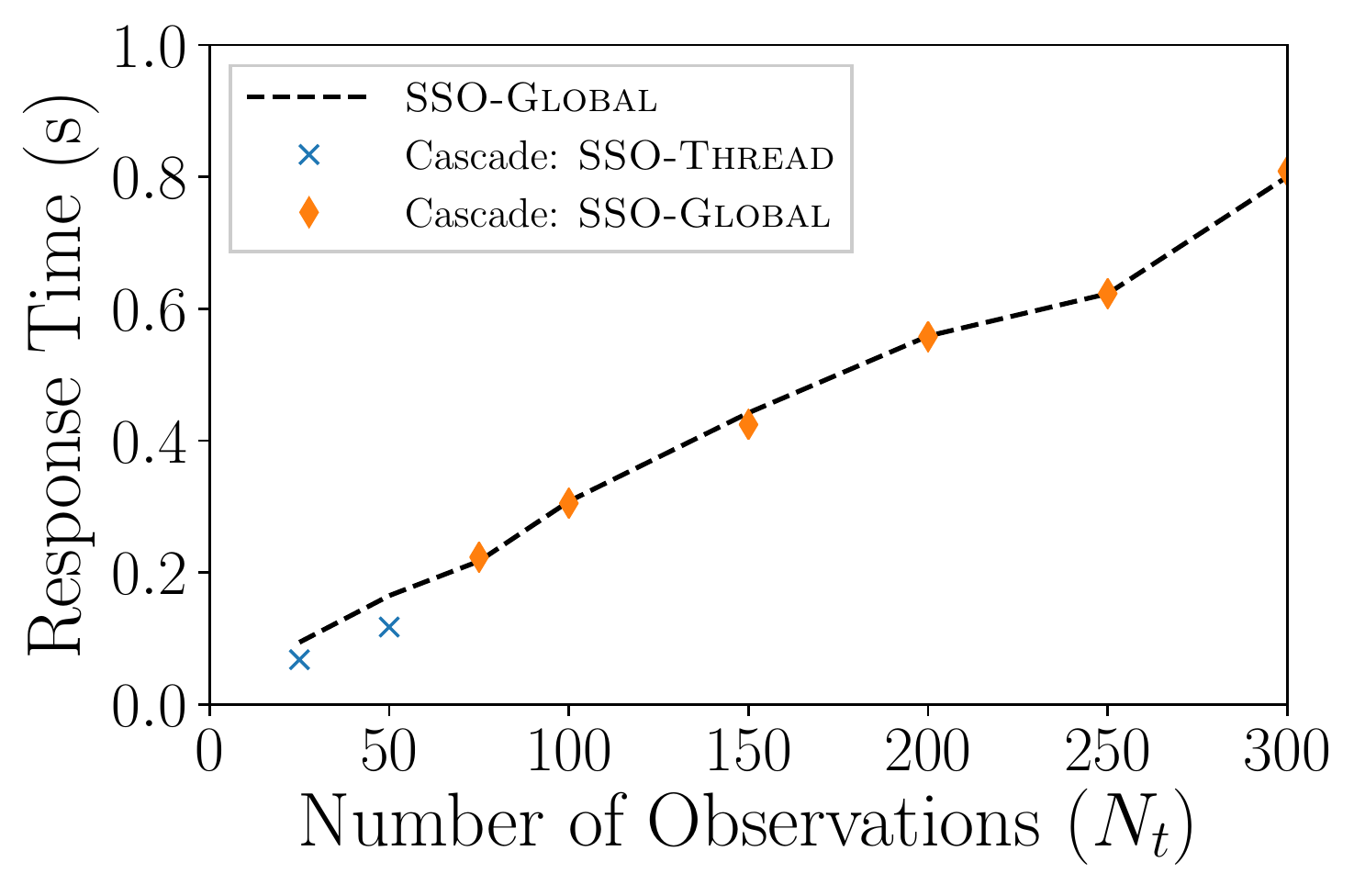}
  \caption{The response time as a function of $N_t$ showing the performance of \ssocascade compared to \ssoglobal when computing the period for a single object (\singleobjnt). The algorithms are executed using a frequency grid of size $N_f=10^5$. The plot shows that the \ssosmthread kernel used by \ssocascade outperforms \ssoglobal when $N_t\leq 50$. When $N_t\geq75$, cascade mode reverts to \ssoglobal due to insufficient shared memory to launch \ssosmthread. Experiments performed on \platforma with FP64.}
   \label{fig:cascade_synthetic_8205}
\end{figure}

As described in Section~\ref{sec:cascade}, in the original \ss algorithm, we attempt to launch the kernel that executes a single thread per frequency (\ssosmthread). This kernel can fail to launch if too much shared memory is requested, which is a function of $N_t$. Consequently, we launch the global memory kernel (\ssoglobal) if \ssosmthread fails to execute. Figure~\ref{fig:cascade_synthetic_8205} plots the response time as a function of $N_t$, where we show the components of \ssocascade (\ssosmthread and \ssoglobal), as the ``x'' and ``diamond'' markers, respectively. For comparison, the dashed black curve shows the execution time of independently launching \ssoglobal. We observe that \ssosmthread is useful for small light curves, as it outperforms \ssoglobal when $N_t\leq 50$. When $N_t\geq 75$, the cascading kernel will execute \ssoglobal as there is insufficient shared memory to launch \ssosmthread. The convergence of the dashed curve and the diamond markers indicates that cascade mode does not degrade performance compared to simply executing \ssoglobal. Thus, when configuring the software, \ssocascade can be used without any performance penalty relative to using the global memory baseline, \ssoglobal.

\subsection{Performance on the Single Object Datasets}
\begin{figure}[!t]
\centering
  \includegraphics[width=1.0\columnwidth]{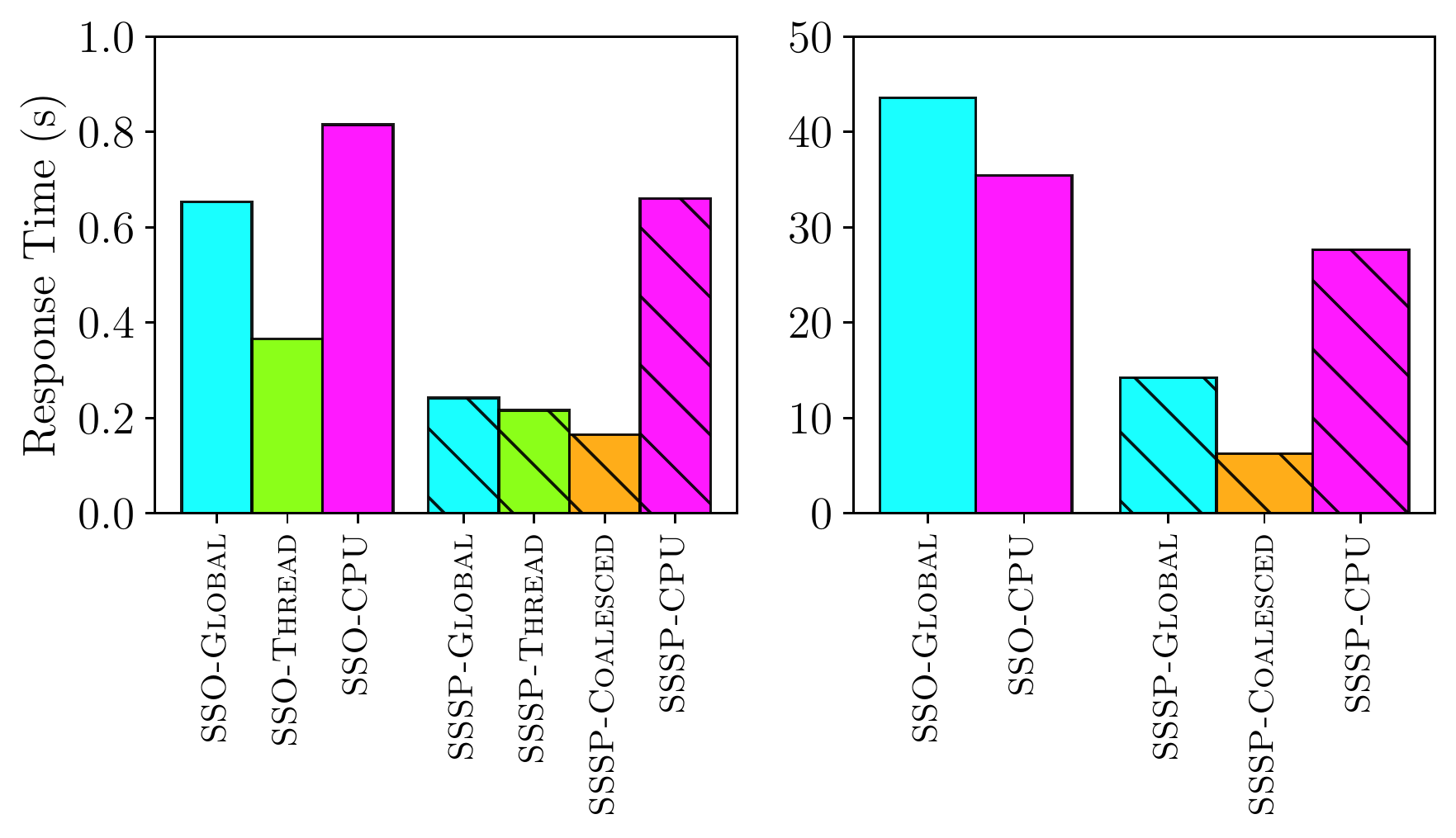}
  \caption{The response time of executing the various kernels on a small and large time series of a single object. \emph{Left:} \singleobjsmall, and \emph{right:} \singleobjlarge. Note that \ssosmthread and \ssspsmthread cannot be executed on \singleobjlarge because there is insufficient shared memory to launch the kernel. On both datasets, frequencies are searched in the range [0.1, 10.0) day$^{-1}$ using $N_f=5\times10^5$. \ss is shown by the plain bars, whereas \sssp is denoted by the striped bars. Experiments performed on \platforma with FP64.}
   \label{fig:single_obj_small_large_LC}
\end{figure}

Figure~\ref{fig:single_obj_small_large_LC} examines the performance of the various GPU kernels on small (left panel) and large (right panel) time series datasets, denoted as \singleobjsmall, and \singleobjlarge. On the \singleobjsmall dataset, \ssosmthread and \ssspsmthread can be executed because there is sufficient shared memory; however, on the \singleobjlarge dataset, there is insufficient memory to launch these kernels. 

From Figure~\ref{fig:single_obj_small_large_LC} (left panel), using the original \ss algorithm on the \singleobjsmall dataset, \ssoglobal performs worse than \ssosmthread. On this dataset, both \ss and \sssp outperform the CPU implementation, \cpusso. However, on the larger dataset, \singleobjlarge  (Figure~\ref{fig:single_obj_small_large_LC}, right panel), \cpusso outperforms \ssoglobal, indicating that the GPU degrades performance relative to using the CPU when executing \ss. 

Comparing the performance of the single pass variant kernels, we find that \ssspcoalesced outperforms the global memory baseline, \ssspglobal and the shared-memory kernel \ssspsmthread on both datasets. Since \ssspcoalesced is not limited by shared memory, and can execute the algorithm on any light curve size, $N_t$, it is preferable to select this GPU kernel over \ssspglobal and \ssspsmthread. We find that \ssspcoalesced outperforms the parallel CPU implementations on both small and large datasets, achieving a speedup of 3.96$\times$ and 4.47$\times$ on the \singleobjsmall and \singleobjlarge datasets, respectively.

\subsection{Kernel Time Breakdown}

To understand the major bottlenecks in the \ss GPU implementations, we profile several kernels using the Nvidia Visual Profiler and report the time spent in major components of the algorithms. While the main \ss kernel is straightforward to measure, the Thrust sorting algorithm calls several independent kernels. Consequently, for clarity, when reporting the time spent sorting on the datasets that only require executing a single batch, we report the duration of time between when the first and last sorting kernel started and ended execution, respectively. In these experiments, since the standard deviation of the time trials is low, we only execute a single time trial.

Table~\ref{tab:kernel_breakdown_singleobj} shows the percentage of time spent on different tasks on the two single object datasets comparing the two kernels for \ss and one kernel for \sssp. Recall from Figure~\ref{fig:single_obj_small_large_LC} that the \ssspcoalesced \sssp kernel performs best, so we do not examine the other kernels. Furthermore,   as shown in the experiment illustrated in Figure~\ref{fig:single_obj_small_large_LC}, the \ssosmthread kernel can be executed on the \singleobjsmall dataset but not the \singleobjlarge dataset due to insufficient shared memory to process the larger light curve. From Table~\ref{tab:kernel_breakdown_singleobj}, we find that on the \singleobjsmall dataset, the \ss kernels, \ssoglobal and \ssosmthread, both require the greatest fraction of time, and the back-to-back sort requires the second largest fraction of time. However, on \sssp, the \ssspcoalesced kernel requires less time than the sorting. Comparing \ssspcoalesced to \ssosmthread and \ssoglobal, we find that the single-pass variant and the associated coalesced memory optimization is able to reduce the response time such that the smoothing function is no longer the bottleneck. Since the sorting function is state-of-the-art from the Thrust library, there is very little that can be optimized to further improve performance of \sssp. We observe similar performance trends on the \singleobjlarge dataset.

\begin{table*}[!t]

\centering
\begin{footnotesize}
\caption{The percentage of time spent on different tasks, comparing the \ss kernels: \ssoglobal, \ssosmthread  and the \sssp kernel: \ssspcoalesced. The ``other'' column refers to all other components not captured by the main tasks, such as: data transfers and memory allocation, kernel execution overhead, and GPU synchronization. The largest percentage is shown in bold face. Percentages may not add up to 1 due to rounding errors. This table highlights times on the single object datasets, \singleobjsmall and \singleobjlarge.} 
\label{tab:kernel_breakdown_singleobj}
\begin{tabularx}{\textwidth}{|X|r|r|r|r|r|} \hline
\multirow{2}{*}{Component (line in Algorithm~\ref{alg:algssoverview})}&\multicolumn{3}{c|}{Dataset: \singleobjsmall}&\multicolumn{2}{c|}{Dataset: \singleobjlarge}\\\cline{2-6}
&\ssoglobal&\ssosmthread&\ssspcoalesced&\ssoglobal&\ssspcoalesced\\\hline
 Main Smoother Kernel (line~\ref{algline:supsmu})&\bf{78.4}&\bf{63.6}&     12.0&\bf{90.0}&     27.2\\
              Sorting (line~\ref{algline:argsort})&     11.0&     19.0&\bf{45.8}&     7.67&\bf{53.8}\\
          Map argsort (line~\ref{algline:mapargsort})&     0.30&     0.60&     6.75&    0.320&     6.25\\
  Compute phased time (line~\ref{algline:periodmodf})&     0.30&     0.60&     1.50&    0.241&     1.70\\
          Free memory&      7.0&     12.7&     30.4&    0.126&    0.766\\
                Other&      3.0&     3.50&     3.55&    1.65&    10.28\\
\hline
\end{tabularx}
\end{footnotesize}
\end{table*}

\begin{table}[!t]
\centering
\begin{footnotesize}
\caption{The same as Table~\ref{tab:kernel_breakdown_singleobj} on the \sdss dataset, demonstrating batch modes on both \ss and \sssp.} 
\label{tab:kernel_breakdown_sdss}
\begin{tabularx}{\columnwidth}{|X|r|r|} \hline
\multirow{2}{*}{Component}&\multicolumn{2}{c|}{Dataset: \sdss}\\\cline{2-3}
&\ssocascade&\ssspcoalesced\\\hline
 Main Smoother Kernel&\bf{53.0}&11.5\\
              Sorting&     13.9&24.0\\
          Map argsort&    0.556&5.47\\
  Compute phased time&    0.546&1.48\\
          Free memory&     10.7&21.1\\
                Other&     21.3&\bf{36.5}\\
\hline
\end{tabularx}
\end{footnotesize}
\end{table}

Table~\ref{tab:kernel_breakdown_sdss} shows the same as Table~\ref{tab:kernel_breakdown_singleobj}, but on the \sdss dataset using batch mode. On the \sdss dataset, the \ssocascade kernel requires a majority of the response time, but sorting, freeing memory and the tasks outlined by ``other'' require a significant amount of work. The \sdss dataset contains 136 RR-Lyrae stars. We find that the \ssocascade mode called the \ssosmthread kernel for 129 objects, and called the \ssoglobal kernel for 7 objects. This further demonstrates that the cascade mode can be highly advantageous by executing the faster  shared-memory kernel on smaller light curves rather than simply relying on the \ssoglobal kernel that is slower but does not have any light curve size limitations.

We also find that compared to the single object experiments in Table~\ref{tab:kernel_breakdown_singleobj}, the ``other'' category in Table~\ref{tab:kernel_breakdown_sdss} requires a significant amount of time when executing \sssp. Because we know that the light curves are small in the \sdss dataset, it is likely that kernel invocations incur significant overhead in this experiment.

In both Tables~\ref{tab:kernel_breakdown_singleobj}~and~\ref{tab:kernel_breakdown_sdss}, we find that freeing memory requires a non-negligible fraction of the total time. When processing batches of objects, we allocate only the memory required to store the data for that object and free the memory when we finish processing the object. While we could over-allocate memory and free it once, this would incur more kernel executions which would increase overhead. Furthermore, since this software will be used in production-grade settings, particularly for community LSST event brokers, we elect to free the memory after use which may prevent future memory leak bugs in our software as it evolves.

Overall, we find that \sssp requires a minority fraction of the overall response time when processing a single object or a batch of objects. Since the other components of the algorithm cannot be optimized, this indicates that our GPU implementation is highly efficient.

\subsection{Processing Catalogs of Objects in Batch Mode}
In this section, we execute \ss on two real-world catalogs of astronomical data, \sdss and \tess.   From Figures~\ref{fig:cascade_synthetic_8205}~and~\ref{fig:single_obj_small_large_LC},  we observe that the original \ss algorithm should use cascade mode, and that the single pass variant should use the global memory kernel with the coalesced memory access optimization. In this section, we execute the algorithms with these optimizations. We execute \ss such that we use a value of $N_f$ that captures the peaks in the periodograms based on typical science cases for these datasets as described in Section~\ref{sec:eval_num_freq}.

\begin{figure}[!t]
\centering
  \includegraphics[width=1.0\columnwidth]{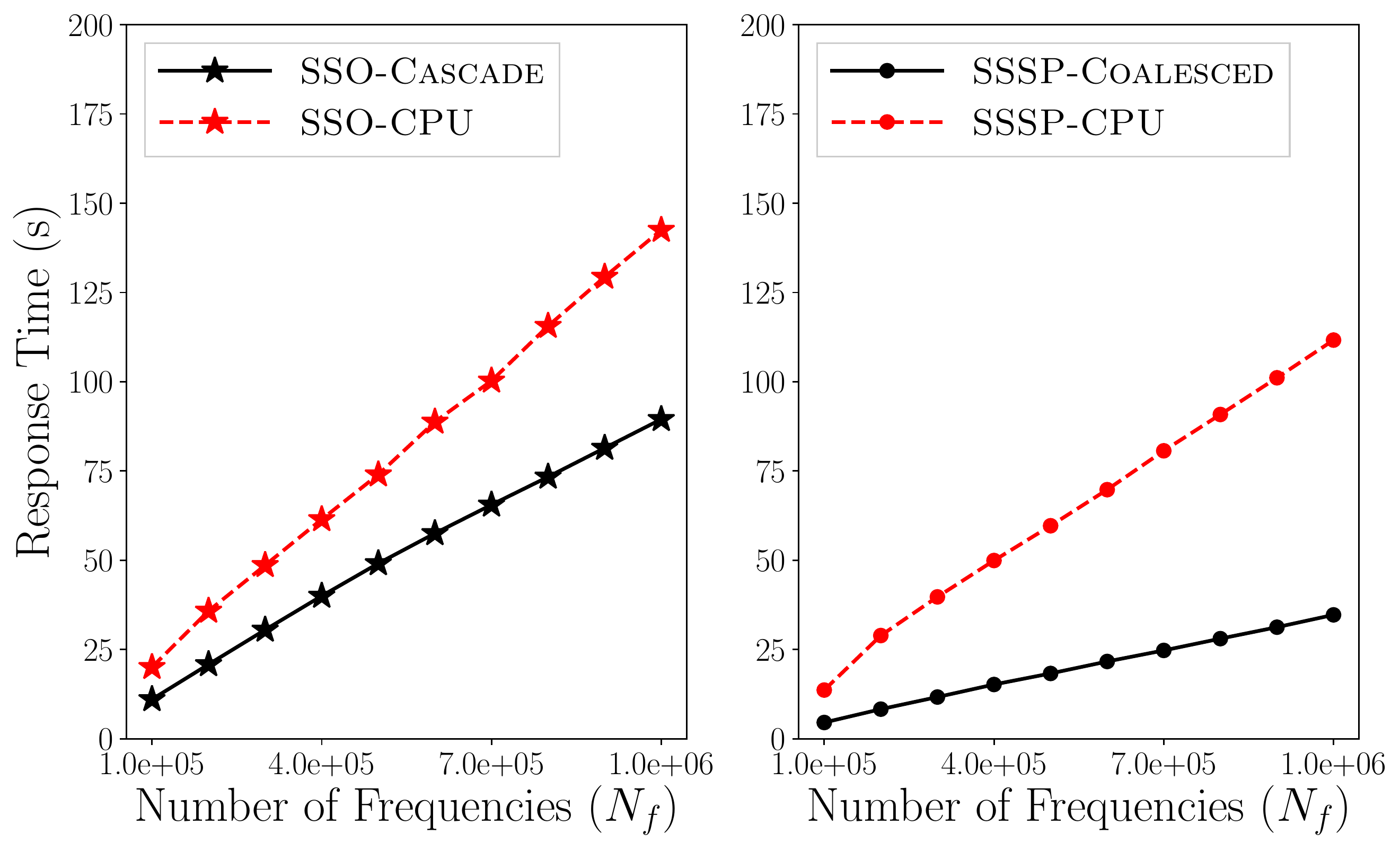}
  \caption{Response time as a function of $N_f$ on the \sdss dataset. \emph{Left:} \ss, and \emph{right:} \sssp.  Frequencies are searched in the range [0.1, 10.0) day$^{-1}$ using $N_f=[10^5, 10^6]$. Experiments performed on \platforma with FP64.}
   \label{fig:batch_sdss}
\end{figure}

Figure~\ref{fig:batch_sdss}  plots the response time for \ss (\emph{Left}) and \sssp (\emph{Right}) as a function of the number of frequencies searched, $N_f$, on the \sdss dataset. We find that when executing \ss, the GPU achieves a speedup over the CPU in the range 1.51--1.82$\times$. On \sssp, the speedup ranges from 3.00--3.49$\times$. The speedup difference between \ss and \sssp is interesting. Because the CPU processes a single frequency at a time, it can store the light curve of an object entirely in L3 cache and then reuse the data multiple times, as the algorithm performs several scans over the data. In contrast, the \sssp algorithm only makes a single scan over the input dataset at each tested frequency; therefore, temporal locality on the CPU is exploited to a lesser degree when using the single-pass variant compared to the original algorithm. Consequently, this results in a larger speedup when using the GPU.

\begin{figure}[!t]
\centering
  \includegraphics[width=1.0\columnwidth]{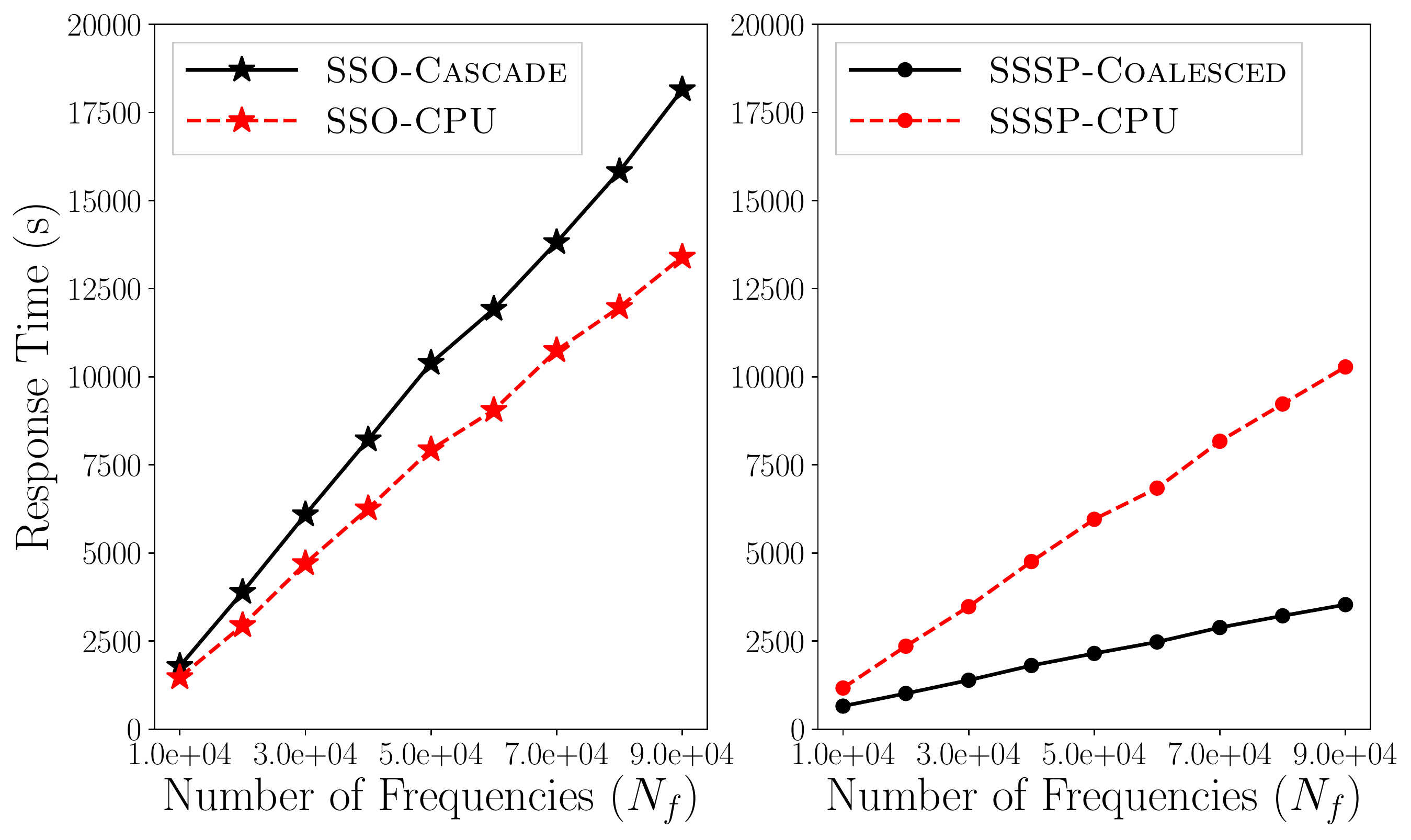}
  \caption{Response time as a function of $N_f$ on the \tess dataset. \emph{Left:} \ss, and \emph{right:} \sssp.  Frequencies are searched in the range [0.1, 10.0) day$^{-1}$ using $N_f=[10^4, 9\times10^4]$. Experiments performed on \platforma with FP64.}
   \label{fig:batch_tess}
\end{figure}

Figure~\ref{fig:batch_tess} plots the response time as a function of $N_f$ on the \tess dataset. We find that when comparing the CPU and GPU \ss algorithms (left panel), the GPU algorithm performs worse than the CPU algorithm, where the GPU yields a slowdown between 0.74--0.82$\times$. In contrast to Figure~\ref{fig:batch_sdss} (left panel) on the \sdss dataset, the number of data points ($N_t$) in each light curve from \tess is greater on average than in \sdss. Consequently, the \ssocascade algorithm requires using the \ssoglobal kernel more frequently than the \ssosmthread kernel, and as was demonstrated in Figure~\ref{fig:single_obj_small_large_LC} (right panel), the CPU outperforms the GPU on large light curves due to high cache reuse that is possible on the CPU.

Figure~\ref{fig:batch_tess} (right panel) illustrates the performance of \sssp on the \tess dataset. We find that the speedup of the GPU over the CPU algorithm is in the range 1.78--2.91$\times$, demonstrating that the single-pass variant is much more efficient on the GPU than the CPU.

We examine the performance of the batch modes as a function of $N_t$ to examine the effect of this parameter on performance. Since the algorithm components do not all have a linear time complexity as a function of $N_t$, such as the sorting step, it is useful to examine how algorithmic performance may degrade with $N_t$. To carry out this experiment, we selected all objects in the \tess dataset with $N_t\geq 500$ which yields 3,150 total objects. We then created input datasets with these objects, where all objects in each dataset have $N_t=50, 100, \ldots, 500$ observations.  

\begin{figure}[!t]
\centering
  \includegraphics[width=1.0\columnwidth]{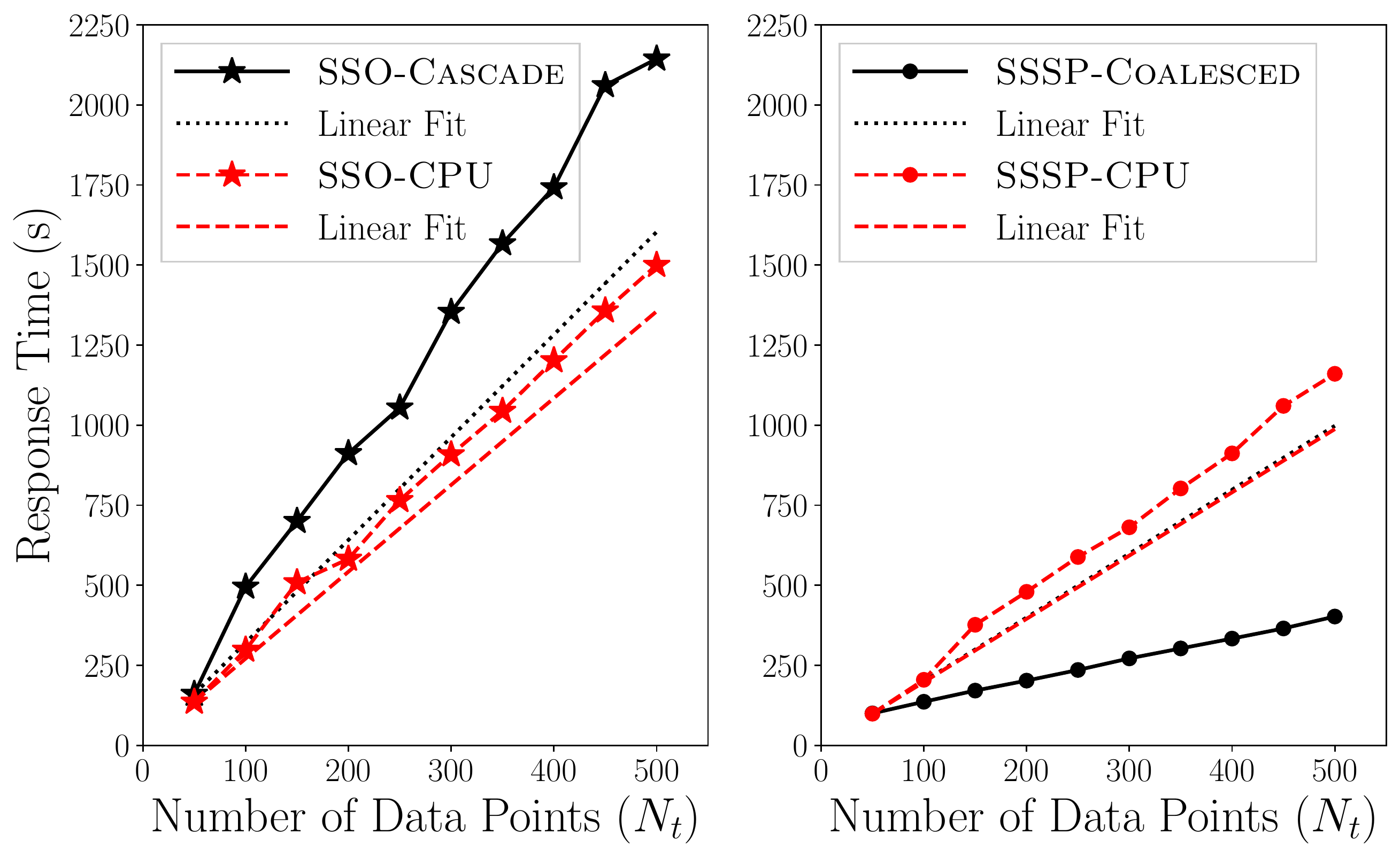}
  \caption{Response time as a function of $N_t$ on the \tess dataset. \emph{Left:} \ss, and \emph{right:} \sssp.  Frequencies are searched in the range [0.1, 10.0) day$^{-1}$ using $N_f=5\times10^4$. The dataset consists of 3,150 objects from the \tess dataset that have $\geq$500 observations per object. The linear fit refers to extrapolating the response time at $N_t=50$ for the CPU and GPU algorithms. Experiments performed on \platforma with FP64.}
   \label{fig:batch_tess_time_vs_num_data_points}
\end{figure}

Figure~\ref{fig:batch_tess_time_vs_num_data_points} plots the response time as a function of the input size ($N_t$) on the \tess dataset for the 3,150 objects described above. Similarly to the results in Figure~\ref{fig:batch_tess}, we find that the GPU performs worse than the CPU algorithm for \ss (Figure~\ref{fig:batch_tess_time_vs_num_data_points}, left panel), whereas the GPU outperforms the CPU on \sssp (Figure~\ref{fig:batch_tess_time_vs_num_data_points}, right panel). The linear fits shown assumes a linear extrapolation as a function of $N_t$, as extrapolated from the response time for the CPU and GPU algorithms at $N_t=50$. We find that the original algorithm degrades with a superlinear profile (Figure~\ref{fig:batch_tess_time_vs_num_data_points}, left panel), and interestingly, the \sssp algorithm exhibits sublinear performance degradation (Figure~\ref{fig:batch_tess_time_vs_num_data_points}, right panel). 

The performance degradation profiles demonstrate why the original algorithm, \ss, is inefficient on the GPU compared to the CPU. The GPU experiences significant superlinear performance degradation, whereas the CPU algorithm degrades gracefully, exhibiting mild superlinear performance degradation. In contrast, the \sssp algorithm exhibits sublinear performance degradation on the GPU, thus scaling to larger numbers of data points without a severe performance penalty, whereas the CPU algorithm exhibits mild superlinear performance degradation.

We omit showing results for the \linear dataset in this section, as \platforma has insufficient main memory to store the periodograms. We examine that dataset in Sections~\ref{sec:eval_fp32_vs_fp64}~and~\ref{sec:eval_multigpu}.

\subsection{Performance and Accuracy of FP32 vs. FP64}\label{sec:eval_fp32_vs_fp64}

\begin{table}[!t]
\centering
\begin{footnotesize}
\caption{The ratio of the response time (s) (FP64/FP32) of \ss and \sssp. The frequency grid uses the same default parameters as shown in Table~\ref{tab:reasonable_nf}. Experiments performed on \platformb using four GPUs. } 
\label{tab:comparison_SS_FP32_vs_FP64}
\begin{tabularx}{\columnwidth}{|L|r|r|} \hline
Dataset&\ssocascade&\ssspcoalesced\\
& Ratio: FP64/FP32& Ratio: FP64/FP32\\\hline
\sdss  &2.92&1.89\\
\tess  &1.25&1.73\\
\linear&1.15&2.05\\
\hline
\end{tabularx}
\end{footnotesize}
\end{table}

\begin{figure*}[!t]
\centering
  \includegraphics[width=1\textwidth]{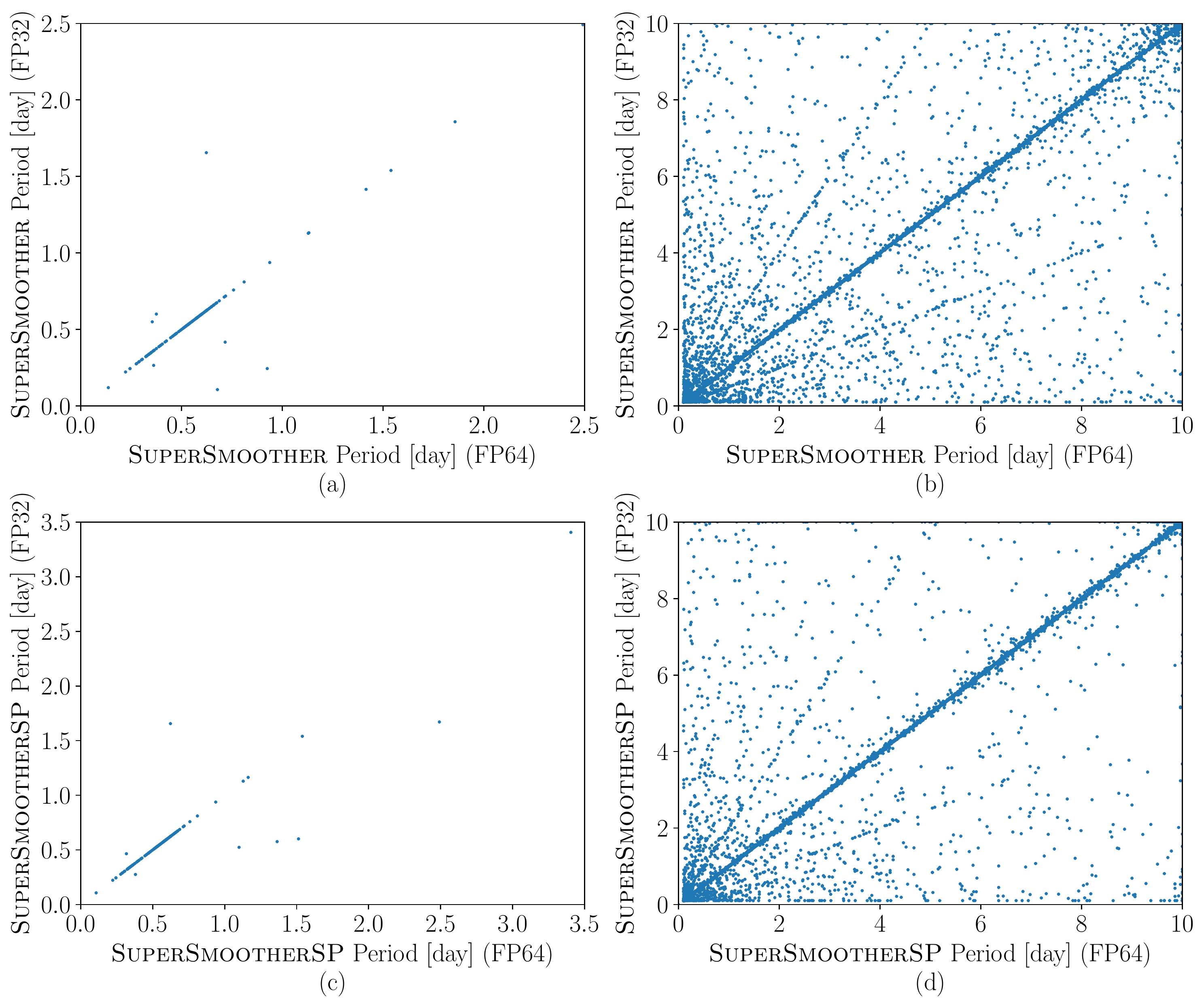}
  \caption{Comparing the periods derived using FP32 and FP64, for \ss on (a) \sdss, and (b) \tess, and \sssp on (c) \sdss, and (d) \tess. Periods that fall on the diagonal line indicate agreement between the FP32 and FP64 solutions. The frequency grid uses the same default parameters as shown in Table~\ref{tab:reasonable_nf}.}
   \label{fig:fp32_vs_64_accuracy}
\end{figure*}

Our software allows the user to select whether they would like to execute the algorithm using FP32 or FP64 precision. In this section, we examine the performance of executing \ss and \sssp with FP32 and FP64, in addition to observing any potential discrepancy in derived solutions. We execute this experiment on \platformb using four GPUs (we will present results for multi-GPU scalability experiments in Section~\ref{sec:eval_multigpu}). Recall from Section~\ref{sec:exp_method} that the GPUs in this platform have limited support for FP64; therefore, the response time ratio of FP64 to FP32 is expected to be high. Consequently, for GPUs with greater support for FP64, the ratios will be lower, and thus, the reported ratios represent an approximate upper bound on the performance degradation yielded by using FP64 over FP32.

Table~\ref{tab:comparison_SS_FP32_vs_FP64} shows the response time ratios of executing \ss and \sssp using FP32 and FP64.  Beginning with \sssp, we find that the response time ratios of FP64/FP32 across the three datasets are in the range 1.73--2.05, indicating that the algorithm has a similar performance penalty across the desperate datasets (e.g., recall that $N_t$ is smaller on average for \sdss than \tess). In contrast, \ss has response time ratios in the range 1.15-2.92 across the three datasets. The shared-memory kernel is exploited to a greater extent on \sdss because $N_t$ is smaller on average compared to \tess and \linear. When executing \sdss with FP32, \ssocascade uses the shared-memory kernel to process all objects, but requires launching the slower global memory kernel to process 7 of 136 objects when using FP64. This explains why using FP64 causes a significant slowdown on \sdss. In contrast, since most objects cannot be processed with the shared-memory kernel on the \tess and \linear datasets, the penalty for using FP64 is much smaller. 

In summary, while FP64 has a larger memory footprint and increases the number of memory accesses compared to using FP32, we do not find that using FP64 yields a factor of 2 increase in the response time compared to FP32 in all instances. The slowdown is data dependent when using \ss, and is much more consistent across datasets with \sssp. To reiterate, since the GPUs in this platform have minimal hardware dedicated to FP64 arithmetic, it is likely that these ratios will be lower on other GPUs, such as those designed for scientific computing.

Figure~\ref{fig:fp32_vs_64_accuracy} plots the derived periods using FP32 compared to FP64 on \sdss and \tess for (a)--(b) \ss  and (c)--(d) \sssp.  Periods along the diagonal line indicate a perfect match. The percentage of periods that match, defined as being within 3\% of each other are as follows: (a) 94.1\%, (b) 93.0\%, (c) 94.9\%, and (d) 95.4\%. From the \tess dataset (Figure~\ref{fig:fp32_vs_64_accuracy}(b)~and~(d)), we clearly observe that some of the mismatched periods are aliases, where the period generated by FP64 is either half or double the period generated when using FP32. This shows that it may be preferable to employ FP64 over FP32 when executing the algorithm. However, FP32 can capture the same period as that derived by FP64 in $\geq$93\% of instances.   

For consistency, our software allows for selecting either FP32 or FP64, and we do not employ mixed floating point precision. Since there may be cases where a user will want to hard code certain tasks with FP32 or FP64, such as using FP64 in the folding step where the time could be unbounded, and using FP32 for the rest of the tasks, we make our source code publicly available so this and other modifications can be carried out.

\subsection{Multi-GPU Scalability}\label{sec:eval_multigpu}
To assess the scalability of our algorithms on multiple GPUs, we use the \linear dataset, which is the largest catalog that we consider in this paper. There is insufficient main memory in \platforma to store the periodograms for each object in the catalog using $N_f=222,000$ (the minimum number of frequencies to search such that we do not miss peaks in the periodogram). Consequently, this experiment is performed on \platformb, as it has 4 GPUs and sufficient main memory.  The \linear dataset contains a large number of objects, and we estimated that to perform three time trials for each experiment would have required roughly a month of computation time. Therefore, in this experiment, we only perform a single time trial for each data point in the plots to limit the time required to run these experiments. However, we note that the standard deviation is very low between time trials\footnote{We executed $N_f=10^5$ frequencies for three time trials using $N_{GPU}=4$ where the time measurements in seconds are as follows: 19,090.2, 19,078.8, 19,078.1, yielding a standard deviation of $\sigma=5.55$.}. Consequently, a single time trial is sufficient for this experiment.

\begin{figure}[!t]
\centering
  \includegraphics[width=1\columnwidth]{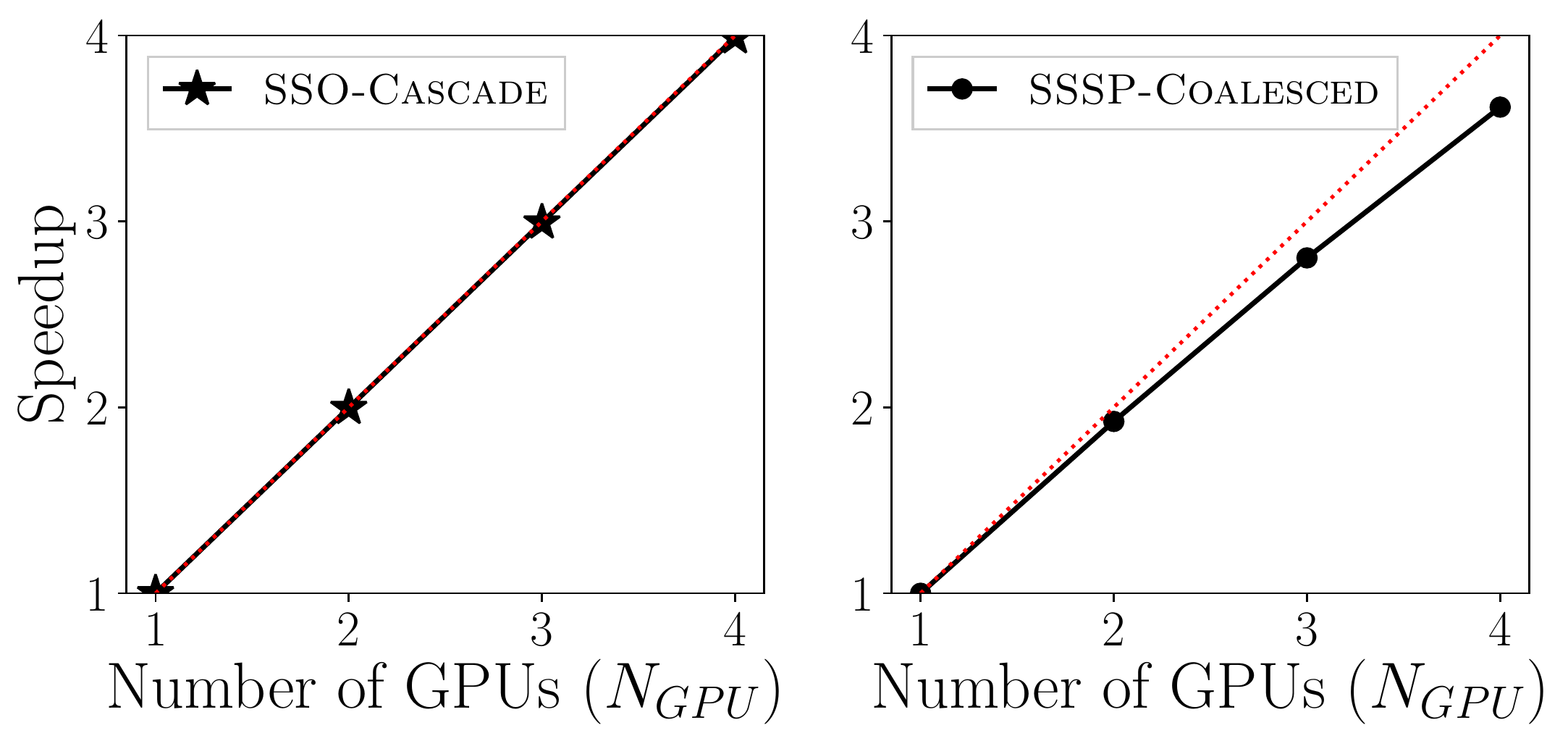}
  \caption{The response time as a function of the number of GPUs $(N_{GPU})$ on the \linear dataset for $N_f=222,000$, and a frequency grid in the range [0.01, 10.0) day$^{-1}$. \emph{Left:} \ssocascade, \emph{Right:} \ssspcoalesced. In each plot, a perfect speedup is indicated by the dashed red line. Experiments performed on \platformb with FP32.}
   \label{fig:multigpu_scalability_linear}
\end{figure}

Figure~\ref{fig:multigpu_scalability_linear} plots the speedup as a function of $N_{GPU}$ on the \linear dataset for \ss (left) and \sssp (right). We find that the multi-GPU \ss implementation achieves a near-perfect speedup. We attribute this to the following factors: (1) The algorithm is compute-bound and there is little contention for PCIe bandwidth between the GPUs; therefore, the host-device interconnect does not limit performance; (2) An object is computed by a single GPU, and the execution time for each object varies with $N_t$. Thus, the probability that two or more GPUs require using the PCIe interconnect at the same time is low. (3) We assign objects to the GPUs using dynamic scheduling in OpenMP. Therefore, there is minimal load imbalance at the end of the computation between GPUs. In summary, multiple GPUs can be employed to significantly reduce the computation time and there is no source of performance degradation due to multi-GPU parallelization. 

On \sssp (Figure~\ref{fig:multigpu_scalability_linear}, right panel), we observe good scalability (a speedup of 3.61$\times$ on 4 GPUs), but the speedup is lower than \ss. This is because the algorithm performs less GPU computation, so there is more contention for host-side resources and the PCIe interconnect.

Comparing the response times used to derive Figure~\ref{fig:multigpu_scalability_linear}, \sssp achieves a speedup of  10.38--11.46$\times$ over \ss.

\subsection{Performance Comparison with Lomb-Scargle}

\begin{table}[!t]
\centering
\begin{footnotesize}
\caption{Response time (s) comparison of the \ss algorithms compared to \ls. The frequency grid uses the same default parameters as shown in Table~\ref{tab:reasonable_nf}. \sdss and \tess are executed on \platforma using FP64, whereas \linear is executed on \platformb using FP32. }
\label{tab:comparison_SS_to_LS}
\begin{tabularx}{\columnwidth}{|X|R|R|R|} \hline
Dataset&\ssocascade&\ssspcoalesced&\ls\\\hline
\sdss  &30.16&11.22&0.27\\
\tess  &5,131.21&1,088.62&160.20\\
\linear&167,731.96&14,634.25&660.84\\
\hline
\end{tabularx}
\end{footnotesize}
\end{table}

In Section~\ref{sec:linear_accuracy_SS_LS}, we showed that on the \linear dataset, \ss and the single-pass variant are less likely than \ls to produce a 24- or 48-hour alias solution. Furthermore, in Section~\ref{sec:singlepass_accuracy_SS_LS}, we showed that the fraction of period matches for \ss and \sssp is higher than for \ls on the \sdss dataset. In this section, we compare \ss to \ls. We use the GPU-accelerated algorithm described in our prior work~\citep{GowanlockAstronomyComputing2021}. The GPU-accelerated \ls algorithm was shown to significantly outperform the parallel CPU algorithm, achieving speedups $>100\times$ on several experimental scenarios~\citep{GowanlockAstronomyComputing2021}.

Table~\ref{tab:comparison_SS_to_LS} shows the response time of deriving the periods using one GPU for the three datasets using the default grid search parameters in Table~\ref{tab:reasonable_nf}.  We find that \ls achieves a speedup over \ss and \sssp in the range of 32.03-253.82$\times$, and 6.80-41.56$\times$, respectively. The \ls algorithm is very well-suited to execution on the GPU. Compared to \ss, it does not have the same pre-processing requirements, and it also has a very small memory footprint.  \ls is useful for period searches where the light curve can be best fit by a sinusoid; however, it is not suitable for light curves of unknown shape. Thus, there is an accuracy vs. performance trade-off between selecting \ls compared to an algorithm that is more robust to non-sinusoidal light curves. 

To address the abovementioned trade-off, in Section~\ref{sec:conclusions} we outline two future research directions, one that takes advantage of the pre-processing step required of \ss and similar methods (described in Section~\ref{sec:introduction}), and the other proposes to combine \ls with \ss to limit the frequency search space.




\section{Discussion \& Conclusions}\label{sec:conclusions}

In this paper, we proposed the first GPU-accelerated \ss algorithm. Furthermore, we proposed a single-pass variant of \ss that uses generalized validation instead of the cross-validation approach used in the original algorithm. We find that the single-pass variant is largely efficient on the GPU, due to requiring fewer scans over the input dataset, and our optimization that reorders data access patterns that exploit coalesced memory accesses to global memory. We also find that the single-pass variant does not provide substantial performance gains over the original algorithm on the CPU due to the high cache reuse that is possible when making several scans over the time series when using the original algorithm. Therefore, for period finding purposes, it may be preferable to use the original algorithm over the single-pass algorithm when using CPU hardware.

\ss and our proposed single-pass variant are expensive algorithms, even with the significant parallelism offered by the GPU's architecture. To reduce the overhead of executing \ss, a future research direction is to employ a \emph{search-and-refine} approach. Here, we will use an inexpensive algorithm, such as Lomb-Scargle, to \emph{search} for potential peaks in the periodogram. Then, we will \emph{refine} the search around the peaks using \ss. This will significantly reduce the cost of searching large frequency spaces with \ss. We can leverage our prior work in this area using our GPU-accelerated Lomb-Scargle algorithm~\citep{GowanlockAstronomyComputing2021}. This method would be similar to~\citet{2004JKAS...37...79S}, which used a coarse-grained period search followed by a fine-grained search.

We found that the pre-processing requirements, such as folding the input time series by a given searched period and then sorting this time series requires non-negligible time (recall that without doing this, the frequencies cannot be searched in parallel). For example, sorting can require the largest fraction of the total time; see Table~\ref{tab:kernel_breakdown_singleobj}. Thus, to leverage this pre-processed data to a greater degree, another future work direction is to execute several period finding algorithms after the data has been pre-processed. This will make better use of the data, as it can be reused across multiple period finders. Using multiple algorithms are likely to yield greater confidence in derived period solutions.

\ss has several performance drawbacks when executed on modern GPU hardware, such as the requirement of sorting the input dataset for each searched frequency. Consequently, another future research direction is to use a hybrid approach that harnesses the capabilities of both the CPU and GPU to concurrently compute the periods of batches of objects. 

\section*{Acknowledgment}
This work has been supported in part by the Arizona Board of Regents, Regents’ Innovation Fund. We thank Will Oldroyd for supplying us with the TESS exoplanet time series data. We thank the referee, Michael Coughlin, for his thorough review and feedback on our manuscript.

\appendix
\section{Open Source Code}\label{sec:code}
The source code is publicly available at~\url{https://github.com/mgowanlock/gpu_supersmoother}. Our code assumes an evenly spaced frequency grid in the range $[f_{min}, f_{max})$. The algorithm uses regular oscillating (not angular) frequencies, where the frequency is given by $f=1/p$, where $p$ is the period. Note that the units output by the code are equivalent to the units in the input dataset file (e.g., typically the time is given as the Julian date). 

The same source code is used for both C and Python interfaces. The C version contains the code used to produce the results in this paper. The Python interface calls the C code using shared libraries. These C shared libraries should be compiled by the user so that they can input their machine-specific parameters via the Makefile. However, we also provide compiled shared libraries if the user does not want to compile the code. To reduce confusion with all of the parameters and optimizations used in the paper, the Python interface selects common default parameters for the user. In particular, in this version of the code, \ss uses cascade mode, and \sssp uses the global memory kernel with the coalesced memory optimization. Additional default parameters are selected and more information can be found in the code repository. The Python interface will take as input the object IDs, time, magnitude, magnitude errors, frequency search ranges, and the selection of \ss algorithm, and will return for each object the object ID, the period with the greatest power for the object, and the periodogram.

\section*{References}


\end{document}